\documentclass[aps,twocolumn,prb,amsmath,amssymb,floatfix,superscriptaddress,reprint,nofootinbib]{revtex4-2} 
\usepackage[normalem]{ulem}
\usepackage[utf8]{inputenc}
\usepackage{graphicx} 
\usepackage{bm}
\usepackage{color}
\usepackage{amsmath,amssymb,amstext,amsbsy,amsthm,bbm}
\usepackage{physics} 
\usepackage{comment}  
\usepackage[dvipsnames]{xcolor} 
\usepackage{cancel}
\usepackage{nicefrac}

\newcommand{\change}[1]{\textcolor{orange}{#1}} 

\newcommand{\quarc}{Quantum Advanced Research Center (QuARC), Consejo Superior de Investigaciones Cient{\'i}ficas (CSIC), 
Madrid, Spain}
\newcommand{\icmm}{Instituto de Ciencia de Materiales de Madrid (ICMM), Consejo Superior de Investigaciones Cient{\'i}ficas (CSIC), 
Madrid, Spain}

\usepackage{braket} 

\makeatletter
\AtBeginDocument{\let\LS@rot\@undefined}
\makeatother

\DeclareMathOperator\Real{Re}
\DeclareMathOperator\Imag{Im}

\usepackage[colorlinks=true,citecolor=Blue, linkcolor=BrickRed,urlcolor=violet]{hyperref}

\begin{document}

	\title{Lindblad theory of linear response susceptibility and dispersive readout in minimal Kitaev junctions}
\author{Tobias Kuhn}
\affiliation{Institute of Physics, University of Augsburg, D-86135 Augsburg, Germany}
\author{Raffael L. Klees}
\affiliation{Institute of Physics, University of Augsburg, D-86135 Augsburg, Germany}

\author{Ram{\'o}n Aguado}
\affiliation{\quarc}\affiliation{\icmm}
\author{M\'onica Benito}
\affiliation{Institute of Physics, University of Augsburg, D-86135 Augsburg, Germany}
\affiliation{Center for Advanced Analytics and Predictive Sciences, University of Augsburg, Augsburg, Germany}

\begin{abstract}
    The field of hybrid superconductor-semiconductor quantum dots is advancing toward the development of functional devices that leverage the advantages of both types of materials. 
    However, the inherent complexity of these devices demands a comprehensive theoretical framework for a complete understanding of
    their responses to external probes, readout and the dissipation arising from environmental coupling. We present a Lindblad-based linear response formalism that captures the
    multi-level nature of these devices, their probe-readout flexibility, and the non-unitary effects of finite-frequency response, including the so-called Sisyphus and Hermes dynamical susceptibilities. These arise from fluctuations in the rates and jump operators, and are hence absent in standard Kubo linear response treatments.  We exemplify the framework using quantum dot-based Kitaev chain setups which are promising candidates for topologically protected Majorana-based parity qubits.
    Our results shed light onto the validity of the standard curvature-based approximation for 
    fermionic parity and qubit readout, show that Hermes terms compensate decoherence in dispersive readout and implement important corrections beyond thermalized states.
\end{abstract}

\maketitle
\section{Introduction}
\label{sec:Introduction}
Superconducting quantum dots (SCQDs) are an exciting prospect for bottom up approaches to parity protected quantum computing \cite{pita-vidal_novel_2025,seoane_souto_subgap_2024}. 
This kind of superconductor-semiconductor hybrid devices feature the controllability of superconductors while leaving a number of microscopic control parameters. An array of SCQDs can form a so-called Kitaev chain \cite{kitaev_unpaired_2001}, which possesses a twofold-degenerate ground-state manifold, whose two states differ by their global fermion parity.
\cite{leumer_exact_2020,sau_realizing_2012,svensson_quantum-dot-based_2024, king_long-range_2025,sanches_revisiting_2026,leijnse_parity_2012,luethi_perfect_2024,samuelson_minimal_2024}. 
In recent years, short versions of Kitaev chains, called minimal Kitaev chains (MKCs), have been realized experimentally in nanowires \cite{dvir_realization_2023,van_loo_single-shot_2026} and two-dimensional electron gases \cite{ten_haaf_two-site_2024, ten_haaf_observation_2025}.
A pair of Kitaev chains forms a Kitaev Josephson junction (KJJ) hosting a fermion-parity qubit~\cite{leijnse_parity_2012,liu_coupling_2024,liu_enhancing_2024,liu_tunable_2022}.
Incorporating this junction 
in a superconducting loop yields a flux-tunable architecture in which a hybrid qubit can be defined
\cite{pino_minimal_2024,christopher_electronically-controlled_2025,pan_rabi_2025,tsintzis_majorana_2024,palacios_effect_2026}. 

Readout of these devices can be performed by means of the quantum capacitance and its flux counterpart, the Josephson inductance. They measure the curvature of the energy bands and are routinely extracted experimentally from the response to weak probe fields~\cite{jennings_probing_2025,lambert_quantum_2016,persson_fast_2010,malinowski_quantum_2022,paila_current-phase_2009,chiodi_probing_2011,baumgartner_josephson_2021}.
Additionally, these measures provide access to the energy spectrum and state populations, enabling the characterization of nonlocal excitations such as Majorana~\cite{trif_dynamic_2018,dourado_assessing_2026}, Andreev~\cite{kurilovich_microwave_2021} and Yu-Shiba-Rusinov \cite{hermansen_inductive_2022} bound states and the readout of fermion-parity qubits~\cite{derakhshan_maman_charge_2020,van_loo_single-shot_2026,liu_quantum_2026,lambert_quantum_2016}.
%
%
Recent theoretical works pointed out that the finite-frequency susceptibility reveals finer details of the underlying physics~\cite{trif_dynamic_2018,park_adiabatic_2020,kurilovich_microwave_2021,peri_unified_2024,hermansen_inductive_2022}.

The response of a system to weak probe fields \cite{kubo_statistical-mechanical_1957} can also be studied while weakly coupled to an environment by means of a 
Lindblad master equation 
to fully describe the response of the dissipative 
dynamics \cite{talkington_linear_2024,peri_unified_2024,wei_linear_2011,ban_linear_2015, campos_venuti_dynamical_2016,albert_geometry_2016,ban_linear_2017,bernazzani_universal_2025,villegas-martinez_application_2016,pan_non-hermitian_2020}. However, previous approaches have either focused on perturbations to coherent dynamics \cite{campos_venuti_dynamical_2016} or on going beyond the linear response regime \cite{shen_non-markovian_2017,boutin_predictive_2025}. While recently the response of the Lindblad-dissipator has been studied as well, providing a closed-form expression for the corresponding response functions remained challenging \cite{levy_response_2021, nie_non-adiabatic_2025,blair_nonequilibrium_2024}. Only recently, analytical solutions have been derived for specific quantum-circuit architectures \cite{peri_unified_2024,kitsenko_reflections_2026}.

\begin{figure}
    \centering
    \includegraphics[width=1\linewidth]{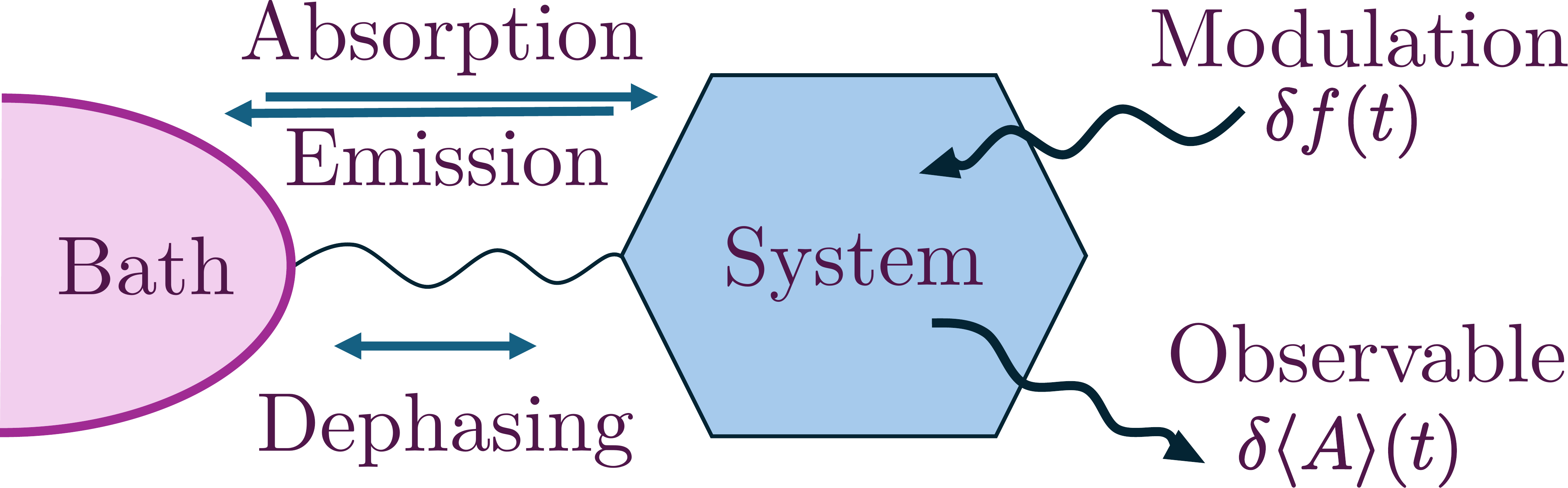}
    \caption{We introduce a parameter fluctuation $\delta f(t)$ into a system that interacts with a Markovian bath. System and environmental dynamics due to absorption, emission and dephasing will be read out in observable $A$. }
    \label{fig:bath-system}
\end{figure}

In this work, we generalize
the Lindblad-based approach to linear response theory introduced in Ref.~\cite{peri_unified_2024} by providing closed form solutions to the finite-frequency susceptibility of an $N$-level quantum system for any probe parameter, observable and reference state. We connect the susceptibility to the band curvature and provide an understanding of the conditions under which curvature-based approximations are justified.
We then analyze the common-gate current susceptibility of a MKC and a KJJ and compare the results with the curvature-based quantum capacitance. 
Finally, we investigate the spectroscopic response to flux perturbations of a KJJ, compare it to the Josephson inductance, and discuss its potential for parity and qubit readout.

The outline of this article is as follows: In Sec.~\ref{sec:Linear Response in open Systems} we present a Lindblad-based linear response theory supplemented by detailed derivations in Appendix~\ref{AppendixA},~\ref{AppendixB}, and~\ref{Appendix:Connection to curvature}. Then, we apply this theory on quantum dot-based Kitaev chains in Sec.~\ref{sec:example_systems}. We present the MKC and compare its dynamical susceptibility to quantum capacitance predictions. Additionally, we couple two minimal chains to form a Kitaev-Josephson junction,
where we highlight dispersive measurements, multilevel effects, and capabilities of flux and gate spectrscopy. Finally, we conclude in Sec.~\ref{sec:Conclusions}.\\


\section{Lindblad-based linear response in open quantum systems}
\label{sec:Linear Response in open Systems}
In this section, we derive closed-form expressions for linear-response functions of an open quantum system.
We consider an $N$-level quantum system described by a Hamiltonian $H = \sum_{m=1}^N E_m \ket{\phi_m} \bra{\phi_m}$ 
with energies $E_m$ and eigenstates $\ket{\phi_m}$.
When such a quantum system is not fully isolated, environmental effects cause decoherence of the quantum states. 
Under Born-Markov approximations\footnote{Response functions have been shown to be quantifiers of non-Markovianity \cite{strasberg_response_2018}, however our derivation will neglect those effects because the bath is assumed to relax quickly. For a Non-Markovian quantum master equation treatment of finite-frequency noise functions see Ref.~\cite{marcos_non-markovian_2011}}, 
namely a weakly coupled environment that has no memory \cite{breuer_theory_2007}, an open quantum system is described by a time-local Lindblad master equation of the form
$\partial_t \rho = \mathcal{L} \rho$,
where $\rho$ is the density matrix of the system. 

The Lindblad superoperator $\mathcal{L}$ extends the von-Neumann equation of the time evolution of a  closed quantum system by additional terms caused by the interaction with the environment.
As briefly outlined in Appendix~\ref{Appendix:System-Bath}, this superoperator takes the form 
\begin{multline} 
    \label{eq:LindbladianGeneralMainText}
    \mathcal{L}\rho = - i [H , \rho] 
    \\
    + 
    \sum_{m,n = 1}^N \Gamma_{mn}\left(L_{mn}\rho L_{mn}^\dag-\frac{1}{2}\big\{L^\dag_{mn}L_{mn},\rho\big\}\right) ,
\end{multline}
where the first term describes the closed-system time evolution, while the second term takes into account additional loss channels with rates $\Gamma_{mn}$ and jump operators $L_{mn} = \ket{\phi_m}\bra{\phi_n}$. 
While $\Gamma_{mn}$ for $m \neq n$ represents a transition from energy level $E_n$ to $E_m$, i.e., absorption ($E_n < E_m$) or emission ($E_n > E_m$) processes, the case $m = n$ with rate $\Gamma_{mm}$ describes pure dephasing affecting the energy level $E_m$.
Finally, the thermal equilibrium state $\rho_\mathrm{th} = \sum_{m=1}^N p_m L_{mm}$ is defined by $\mathcal{L}\rho_\mathrm{th}=0$ and its populations $p_m$ are solely determined by the environmental transition rates $\Gamma_{mn}$ ($m \neq n$) such that detailed balance $\Gamma_{mn} p_n = \Gamma_{nm} p_m$ holds.

To experimentally probe a quantum system, a given parameter $f$ is weakly modulated in time, $f\to f+\delta f(t)$. 
In general, both $\rho \to \rho + \delta \rho(t)$ and all components of $\mathcal{L} \to \mathcal{L} +  \delta\mathcal{L}(t)$ may depend on this parameter and inherit the modulation to first order.
From the master equation $\partial_t \rho = \mathcal{L} \rho$ we can identify all terms that are first-order in the correction, namely
\begin{equation}
    \label{eq:correctionToDensityMatrix}
    \partial_t\,\delta\rho(t)
    =
    \mathcal{L}\,\delta\rho(t)
    + 
    \delta\mathcal{L}(t) \, \rho,
\end{equation}
which is an inhomogeneous ordinary differential equation for the correction $\delta\rho(t)$, where the inhomogeneity explicitly depends on the reference state $\rho$ of the system.

In linear response, we are interested in the corrections to the expectation value $\braket{A} = \mathrm{tr}(A\rho)$ of an arbitrary system observable $A$.
As $A$ can generally also depend on the modulated parameter $f$, we also assume that $A\to A + \delta A(t)$.
As detailed in Appendix~\ref{Appendix:expectation_value}, the first-order correction $\delta\langle A \rangle(t) = \mathrm{tr}(\delta A(t) \rho) + \mathrm{tr}(A \, \delta\rho(t))$ can then be expressed as the Lindblad-based Kubo formula \cite{albert_geometry_2016,wei_linear_2011,ban_linear_2015,campos_venuti_dynamical_2016,villegas-martinez_application_2016} 
\begin{equation}
	\label{eq:ulr_Kubo-formula}
	\delta\langle A \rangle (t) = \int^{\infty}_{-\infty} \chi(t-\tau)\,\delta f(\tau) \, d\tau,
\end{equation} 
with the linear-response susceptibility 
\begin{align}
\chi(t)
&=
\tr{A^\prime\rho}\,\delta(t) 
+
\tr{ A e^{t \mathcal{L}}\mathcal{L}^\prime\rho}\,\Theta(t) \label{eq:ulr_susceptibility_general}\,.
\end{align}
The equation is time-local given a time independent reference state, $\rho\neq \rho(t)$.
The prime notation denotes the derivative with respect to the modulated parameter $f$, e.g., $A'=\partial_fA$, so that $\delta \mathcal{L}(t) = \mathcal{L}' \delta f(t)$ and $\delta A(t) = A' \delta f(t)$.

We name the first term in Eq.~\eqref{eq:ulr_susceptibility_general} static susceptibility, 
\begin{align}
    \chi_\mathrm{st}(t) = \tr{A^\prime\rho}\,\delta(t) .
    \label{eq:static_susceptibility}
\end{align}
It describes the response due to the change of $A$ by modulation of $f$.
In the second term of Eq.~\eqref{eq:ulr_susceptibility_general}, we see that dynamical contributions appear due to changes of the Lindblad superoperator. 
Those changes $\mathcal{L}'$ are generated by modulation-induced changes in the Hamiltonian $H$, the rates $\Gamma_{mn}$, and the jump operators $L_{mn}$ \cite{peri_unified_2024}. 
In this way, we can write $\mathcal{L}' = \mathcal{L}_H' + \mathcal{L}_\Gamma' +\mathcal{L}_L'$, with a Hamiltonian contribution $\mathcal{L}_H'$ and the two so-called Sisyphus and Hermes contributions $\mathcal{L}_\Gamma'$ and $\mathcal{L}_L'$, respectively.
Physically, they correspond to coherent level mixing, modulation in bath-induced transition rates and changes of the dissipative channels themselves.
This separation allows us to split the total susceptibility $\chi(t)$ into four contributions,
\begin{align}
    \label{eq:total_susceptibility}
    \chi(t) = \chi_\mathrm{st}(t) + \chi_{H}(t) + \chi_{\Gamma}(t) + \chi_{L}(t),
\end{align}

where
\begin{align}
\label{eq:other_contributions_susceptibility}
    \chi_{\alpha}(t) = 
    \tr{ A e^{t \mathcal{L}}\mathcal{L}_\alpha' \rho}\Theta(t) 
\end{align}
for $\alpha \in \{H,\Gamma,L\}$.
In the following, we discuss in detail the different contributions in the frequency domain,
\begin{align}
    \chi(\omega) = \int_{-\infty}^\infty \chi(t) e^{i\omega t} dt ,
\end{align}
which we derive for general diagonal mixed reference states $\rho = \sum_{m=1}^N p_m L_{mm}$ with populations $p_m$ that do not necessarily follow a thermal distribution.
As pointed out in Ref.~\cite{peri_unified_2024}, it is possible to reframe this finite frequency response in terms of the admittance and the linear  elements of an electrical circuit; see Appendix~\ref{Appendix:Admittance for gate perturbations}.

\subsection{Hamiltonian contribution}
The Hamiltonian contribution $\chi_{H}(\omega)$ originates from first-order corrections to the Hamiltonian, $H \to H + H' \delta f(t)$, generating the first-order change of the Lindblad superoperator
\begin{equation}
	\mathcal{L}_H' \rho = -i [H', \rho] .
\end{equation} 
It governs coherent transition dynamics induced by the modulation. As derived in Appendix 
\ref{Appendix:derivation_Hamiltonian}, the Hamiltonian susceptibility can be written as 
\begin{align}
    \label{eq:ULR_susceptibility-H}
    \chi_H(\omega) 
    = 
    \sum_{\substack{m,n=1\\(m\neq n)}}^N
    (p_m-p_n) 
    \frac{ \braket{\phi_m|H'|\phi_n}
    \braket{\phi_n|A|\phi_m} }{E_m-E_n-\omega -i\Gamma_{T_2}^{mn}} 
\end{align} 
in terms of the transition energies $E_m-E_n$ between eigenstates $\ket{\phi_m}$ and $\ket{\phi_n}$ of the unperturbed Hamiltonian $H$ and the decoherence rates 
\begin{align}
    \label{eq:general_Gamma_T_2_maintext}
    \Gamma_{T_2}^{mn} = \frac{1}{2}\sum_{k=1}^N (\Gamma_{km} +\Gamma_{kn})
\end{align}
of the off-diagonal elements of the density matrix.
Together with the static susceptibility and for $A=H'$, this 
contribution reproduces 
results that are already known in literature \cite{park_adiabatic_2020}.

\subsection{Sisyphus contribution}
In quantum dot setups, the Sisyphus term has been shown to be important to understand dynamical dissipation \cite{peri_unified_2024,persson_excess_2010,esterli_small-signal_2019,peri_beyond-adiabatic_2024}.

The Sisyphus contribution $\chi_\Gamma(\omega)$ originates from first-order corrections to the rates, $\Gamma_{mn} \to \Gamma_{mn} + \Gamma_{mn}' \delta f(t)$, generating the first-order change of the Lindblad superoperator
\begin{equation}
\label{eq:perturbed_Lindbladian_sisyphus}
	\mathcal{L}_\Gamma' \rho 
    =  
    \sum_{m,n = 1}^N \Gamma_{mn}' \left(L_{mn} \rho L_{mn}^\dag - \frac{1}{2}\big\{L^\dag_{mn}L_{mn},\rho \big\}\right) .
\end{equation} 
As shown in Appendix~\ref{Appendix:derivation_Sisyphus_corrected},
the Sisyphus susceptibility can be written as 
\begin{align}
    \label{eq:Sisyphus solution}
    \chi_\Gamma(\omega) 
    &= 
    \sum_{\substack{k,l,m,n=1\\(m\neq n,\lambda_k < 0)}}^N
    c_{k,lmn} \, \Gamma_{mn}' \, p_n \, 
    \frac{ i  \braket{\phi_l|A|\phi_l}  }{\omega-i\lambda_k} 
    ,
\end{align}
with expansion coefficients
\begin{align}
    c_{k,lmn}
    =
    \braket{\phi_l|r_k|\phi_l}
    \left( 
    \braket{\phi_m|l_k|\phi_m} 
    -
    \braket{\phi_n|l_k|\phi_n} 
    \right) .
\end{align}
Here, $r_k$ and $l_k$ are the right and left eigenoperators of the Lindblad superoperator, respectively, in the subspace spanned by the dephasing jump operators $L_{mm}$.
They satisfy the eigenvalue equations $\mathcal{L} r_k = \lambda_k r_k$ and $\mathcal{L}^\dag l_k = \lambda_k l_k$, where $\lambda_k \leq 0$ are $N$ real eigenvalues; see Appendix~\ref{section:diagonalizationSuperoperatorAppendix} for details.

In general, there is a unique zero eigenvalue for which the right eigenoperator is the thermal equilibrium density matrix $\rho_\mathrm{th}$ with $\mathcal{L}\rho_\mathrm{th} = 0$, as discussed already in the beginning of Sec.~\ref{sec:Linear Response in open Systems}, while the corresponding left eigenoperator is the identity.
For this thermal state, the expansion coefficient is zero and, hence, it is excluded from the summation in Eq.~\eqref{eq:Sisyphus solution}.
For general $N$-level systems, the remaining nonzero coefficients $c_{k,lmn}$ for the other $N-1$ nonzero eigenvalues are cumbersome expressions that have to be solved for numerically.
For $N=2$ states, however, the nonzero coefficients $c_{k,lmn}$ become quite simple (see Eq.~\eqref{eq:sisyphus_coefficient_TLS} in Appendix~\ref{Appendix:derivation_Sisyphus_corrected}), resulting in the Sisyphus susceptibility
\begin{align} 
    \label{eq:sysiphusTLS}
    \chi_\Gamma(\omega) 
    &=  
    ( \Gamma_{12}' \, p_2 - \Gamma_{21}' \, p_1 ) 
    \frac{ \braket{\phi_1|A|\phi_1} - \braket{\phi_2|A|\phi_2} }{\Gamma_{12} + \Gamma_{21} - i \omega} 
    ,
\end{align}
which recovers the result for two-level systems in Ref.~\cite{peri_unified_2024}.

In addition for $A = H'$, it was shown in Ref.~\cite{peri_unified_2024} that Sisyphus contributions lead to additional linear conductance and capacitance terms in gate spectroscopy when translated to quantum electric circuits.

Interestingly, $\chi_\Gamma(\omega)$ is scaling only with changes in absorption and emission rates by the modulated parameter, respectively, while pure dephasing has no effect on $\chi_\Gamma(\omega)$. 
Furthermore, Sisyphus terms will be negligible in systems with flat energy bands with respect to the 
modulated parameter $f$ resulting in small $\Gamma_{mn}'$.

Eqs.~\eqref{eq:Sisyphus_DBBT} in Appendix~\ref{Appendix:derivation_Sisyphus_corrected} show that it is possible to separate $\chi_\Gamma$ into a detailed balance enforced term $\chi_{\Gamma,\mathrm{DB}}$ and its corrections beyond thermalization $\chi_{\Gamma,\mathrm{BT}}$. Remarkably, the detailed balance contribution $\chi_{\Gamma,\mathrm{DB}}$ in the Sisyphus susceptibility language can be rewritten solely in terms of population derivatives, rather than rate derivatives. This reveals that $\chi_{\Gamma,\mathrm{DB}}$ is the Lindblad-based analog to the diagonal susceptibility derived in Ref.~\cite{trif_dynamic_2018} and references therein. On the other hand, deviations from detailed balance captured by $\chi_{\Gamma,\mathrm{BT}}$ cannot be captured by previous approaches.

\subsection{Hermes contribution}
In the development of unified linear response theory \cite{peri_unified_2024}, an additional term has been introduced that bridges the gap between semiclassical and input-output theory. This so-called Hermes admittance emerges as a natural consequence of the Lindblad formulation. As such, it takes into account that environmental transitions occur between modulated eigenstates rather than a fixed basis.

In general, the Hermes contribution $\chi_L(\omega)$ originates from first-order corrections to the jump operators $L_{mn} \to L_{mn} + L_{mn}' \delta f(t)$ due to changes in the eigenstates 
$\ket{\phi_m} \to \ket{\phi_m} + \ket{\phi_m'} \delta f(t)$.
This results in the first-order change of the Lindblad superoperator 
\begin{multline}
\label{eq:dissipator in linear response}
    \mathcal{L}_L'\rho
    =
    \sum_{m,n = 1}^N \Gamma_{mn} 
    \left( L_{mn}' \rho L_{mn}^\dag
    +
    L_{mn}\rho (L_{mn}^\dag)'
    \right.
    \\
    -
    \frac{1}{2} \left. \left\{
    (L_{mn}^\dag)' L_{mn}
    +
    L_{mn}^\dag
    L_{mn}'
    ,\rho 
    \right\}
    \right).
\end{multline} 
As shown in Appendix~\ref{Appendix:derivation_Hermes}, the Hermes susceptibility can be written as 
\begin{align}
    \label{eq:HermesContributionMainText}
    \chi_L(\omega) 
    &=   
    \sum_{\substack{m,n=1\\(m\neq n)}}^N 
    \frac{i \Lambda_{mn}}{E_m - E_n}
    \, 
    \frac{  \braket{\phi_m|H'|\phi_n} \braket{\phi_n|A|\phi_m} }{ E_m - E_n - \omega -i \Gamma_{T_2}^{mn} } ,
\end{align}
where $\Lambda_{mn} = - (p_m - p_n ) \Gamma_{T_2}^{mn} + \mathcal{G}_m - \mathcal{G}_n$ and $\mathcal{G}_m = \sum_{k=1}^N ( \Gamma_{km} p_m - \Gamma_{mk} p_k )$.
Here, $\mathcal{G}_m$ is a measure of the thermalization of the system. 
For a completely thermalized state $\rho_\mathrm{th}$, detailed balance leads to $\mathcal{G}_m = 0$ and, thus, $\chi_L(\omega)$ is only influenced by the decoherence rates $\Gamma_{T_2}^{mn}$ in Eq.~\eqref{eq:general_Gamma_T_2_maintext}. 
The introduction of $\mathcal{G}_m$ is an important novelty of this paper as previous approaches \cite{trif_dynamic_2018,peri_unified_2024} only considered  
populations in thermal equilibrium. 
Therefore, we will later in Sec.~\ref{sec:example_systems} separate $\chi_L=\chi_{L,\mathrm{DB}}+\chi_{L,\mathrm{BT}}$  into detailed balance $\chi_{L,\mathrm{DB}}$ and beyond thermalization $\chi_{L,\mathrm{BT}}$ contributions, the latter featuring only $\mathcal{G}_m$ terms, as defined in Eqs.~\eqref{eq:chiLDB_BT}.

The Hermes susceptibility $\chi_{L,\mathrm{DB}}$ is part of the Lindblad-based analog to the non-diagonal susceptibility derived in Ref.~\cite{trif_dynamic_2018} and references therein\footnote{Using the Kubo formalism, Ref.~\cite{trif_dynamic_2018} employs quasi-equilibrium rate equations 
$$\dot{\rho}(t)=-i[H(t),\rho(t)]-\Gamma \left(\rho(t)-\rho_{eq}(t)\right)$$
with relaxation superoperator $\Gamma$ and instantaneous equilibrium density matrix $\rho_{eq}(t)$. The susceptibility is then decomposed into curvature, diagonal and non-diagonal terms. This approach assumes decoupled equations for diagonal and off-diagonal quasi-equilibrium density matrix elements as well as constant rates within $\Gamma$. For a Lindblad formalism, only the non-diagonal quasi-equilibrium dynamics decouple from each other and rates can vary.
Thus, only $\chi_{L,\mathrm{DB}}$ is reproduced exactly. Other terms are beyond a simple comparison.}. 
As shown below, this additional response is essential for understanding why dispersive measurements remain remarkably robust against decoherence.

\subsection{Relation between dispersive susceptibility and band curvature}

In the case of 
$A=H'$, the Hamiltonian susceptibility in Eq.~\eqref{eq:ULR_susceptibility-H} represents a population-modified
second-order perturbation theory expression broadened by dissipation. 
As shown in Appendix~\ref{sec:lowFreqSpecIsoSysAppendix} for negligible environmental interactions, $\Gamma_{mn} = 0$, the zero-frequency susceptibility collapses to a weighted sum of energy level curvatures, 
\begin{equation}
	\chi(0) = \chi_\mathrm{st}+\chi_H(0) = \sum_{m=1}^N p_m E_m'' .
    \label{eq:curvature}
\end{equation}

This recovers the known  quantum capacitance $C_Q$ in the case of gate spectroscopy when the parameter $f = \mu$ is a chemical potential of the system
\cite{van_loo_single-shot_2026,boutin_predictive_2025,microsoft_azure_quantum_interferometric_2025,liu_quantum_2026,zhang_gate_2025}.
In this case, the total susceptibility including dissipation and finite-frequency contributions implies additional conductive and capacitive components to the isolated counterpart, the quantum capacitance. 
When the parameter $f=\phi$ is the phase difference between two superconductors, then $\chi(0)=L^{-1}_J$ recovers the inverse Josephson inductance \cite{paila_current-phase_2009,trif_dynamic_2018}.

In comparison with the Hamiltonian susceptibility $\chi_H(\omega)$ in Eq.~\eqref{eq:ULR_susceptibility-H}, the Hermes contribution $\chi_L(\omega)$ looks similar up to a different factor $i\Lambda_{mn}/(E_m-E_n)$ and is, therefore, smaller in well-isolated systems and for large energy gaps.
For $A=H'$, this similarity causes Hermes contributions $\chi_{L,\mathrm{DB}}$ to compensate the decoherence effects that suppress the Hamiltonian term and, thus, forces the weighted sum of energy level curvatures to reemerge in the dispersive limit.
One can show that
\begin{align}
    \label{eq:dispersiveLimitHamiltonianAndHermesMainText}
    \chi_\mathrm{st} 
    + \chi_H(0)  
    + \chi_{L,\mathrm{DB}}(0)
    =   
    \sum_{m=1}^N p_m E_m''
    ,
\end{align}
which is independent of the environmental decoherence rates $\Gamma_{T_2}^{mn}$. This effect is caused by the Hermes contribution $\chi_{L,\mathrm{DB}}$ to maintain quantum coherence and has been mentioned before in Ref.~\cite{kitsenko_reflections_2026}.
The full result including corrections $\chi_{L,\mathrm{BT}}$ beyond thermalization, i.e., $\mathcal{G}_m \neq 0$, which also explicitly depends on the decoherence rates $\Gamma_{T_2}^{mn}$, is provided in Appendix~\ref{sec:dispersive_limit}. The corrections are always real-valued and only negligible if \begin{equation}
    \left|\frac{\mathcal{G}_m-\mathcal{G}_n}{E_m-E_n}\right|\ll 1\,.
\end{equation}
Thus, this condition gives an important threshold whether dispersive measurements are protected from decoherence.

Finally, the curvature limit can also be found for systems in which the dominant contribution to decoherence originates from dephasing, $\Gamma_{mm} \gg \Gamma_{mn}$ $(m\neq n)$, as shown in Appendix~\ref{sec:dephasing_limit}.
In the limit of strong pure dephasing ($\Gamma_{mn} = 0$ for $m\neq n$), we also get the dispersive result
\begin{align}
    \label{eq:dephasing_limit}
    \lim_{\Gamma_{mm} \to \infty}\chi(\omega) = \sum_{m=1}^N p_m E_m'',
\end{align}
since both $\mathcal{G}_m = 0$ and $\chi_\Gamma(\omega) = 0$.
This was already deduced in Ref.~\cite{peri_unified_2024}, where it was interpreted as 
the semiclassical limit.

Therefore, dispersive probes are generally unaffected by pure dephasing while relaxation effects are only suppressed for completely thermalized systems.

\section{Finite frequency susceptibility of Kitaev chain systems}
\label{sec:example_systems}
 To demonstrate the capabilities of the Lindblad-based linear response theory, we investigate semiconductor-superconductor hybrid systems.
In particular, a minimal Kitaev chain via proximitized quantum dots \cite{leijnse_parity_2012,leumer_exact_2020,ten_haaf_two-site_2024,dourado_majorana_2025,ten_haaf_observation_2025, van_loo_single-shot_2026} and a
superconducting junction of two minimal Kitaev chains.  The former is a promising candidate to study and control Majorana zero modes in bottom up quantum dot arrays. 
The latter supports the so-called Majorana (or parity) qubit \cite{leijnse_parity_2012,aguado_majorana_2020,tsintzis_majorana_2024,pan_rabi_2025} and a 
recently proposed hybrid qubit when embedded in a transmon circuit, the Kitmon qubit \cite{pino_minimal_2024}.

\subsection{Minimal Kitaev Chain (MKC)}
\begin{figure*}[t]
	\centering
		\includegraphics[width=\textwidth]{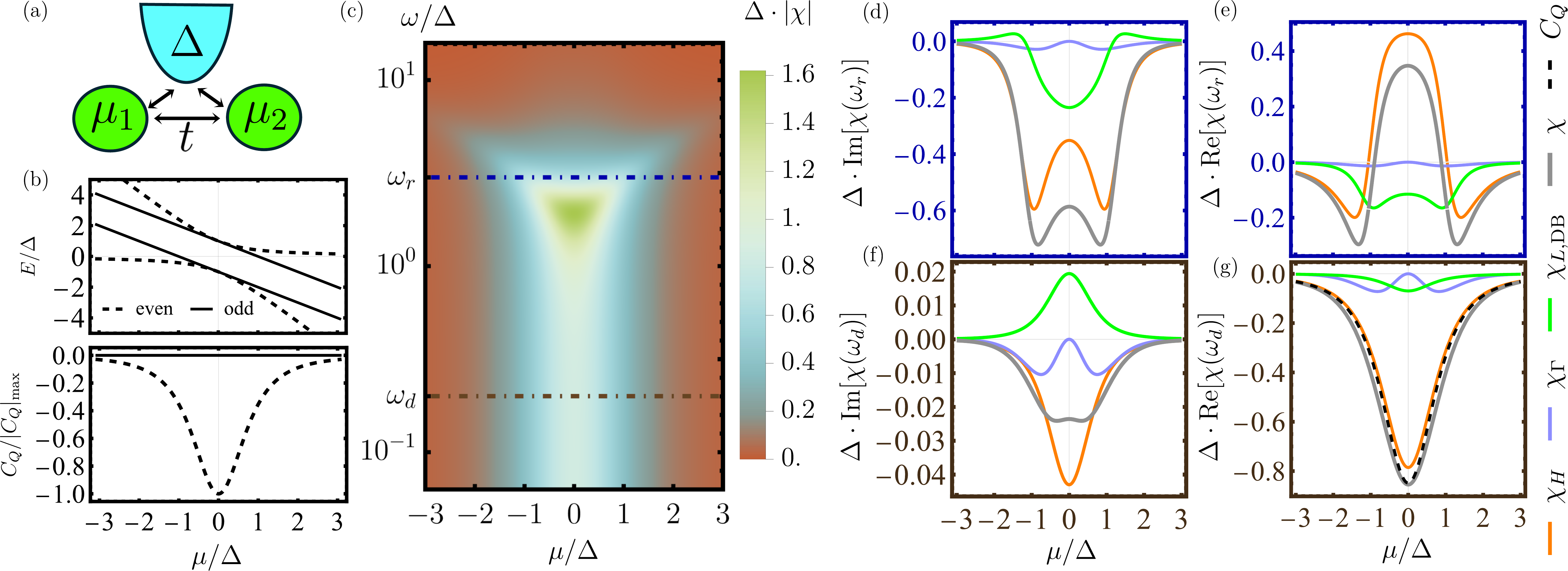}

	\hfill
	\caption{
    Common gate spectroscopy on a minimal Kitaev chain. 
    \textbf{(a)} Two quantum dots with onsite energies $\mu_1$ and $\mu_2$ are proximitized by a superconductor. Induced crossed Andreev reflection $\Delta$ and direct elastic cotunneling $t$ mediate the hopping between the dots.
    \textbf{(b)} 
    Energy spectrum $E$ and quantum capacitance $C_{Q}$ (normalized by its maximal value $|C_{Q}|_\mathrm{max}$) in the odd- and even-parity subspaces as a function of the average onsite energy $\mu = (\mu_1+\mu_2)/2$.
    \textbf{(c)} Modulus of the total susceptibility $|\chi|$ from Eq.~\eqref{eq:total_susceptibility} for different probe frequencies $\omega$ and average onsite energies $\mu$. 
    A resonant regime at $\omega \approx \omega_r$ and a dispersive regime at $\omega \approx \omega_d$ emerge. 
    \textbf{(d-g)} Imaginary and real part of the total susceptibility $\chi$ and its contributions $\chi_H$, $\chi_\Gamma$, $\chi_L$, evaluated along the slices \textbf{(d,e)} $\omega = \omega_r$  and \textbf{(f,g)} $\omega = \omega_d$ in panel \textbf{(c)}. We show the quantum capacitance $C_Q$ in panel \textbf{(g)} for comparison.
    Parameters: $\mu_1=\mu_2$,
    $k_BT=4\Delta/5$, 
    $|s_{\pm}|^2=1/2$, 
    $\Gamma_\phi=0$,
    $t=\Delta$, 
    $\omega_d=\Delta/5$,
    $\omega_r=3\Delta$.}
	\label{fig:min_chain_susc}
\end{figure*}

The minimal realization of a Kitaev chain consists of two quantum dots with onsite energies $\mu_j$ proximitized by a superconductor,
which is sketched in Fig.~\ref{fig:min_chain_susc}a.
The effective Hamiltonian of the MKC is defined as
\begin{align}
	H=-\sum_{j=1}^{2}\mu_j d_j^\dagger d_j+\Delta d_1^\dagger d_{2}^\dagger-td_1^\dagger d_{2}+ \mathrm{h.c.} ,
    \label{eq:min_chain_Hamil}
\end{align}
where hopping between sites is mediated by crossed Andreev reflection (CAR) $\Delta$ and elastic cotunneling (ECT) $t$. $d_j$ and $d_j^\dag$ are spinless fermionic annihilation and creation operators on quantum dot $j = 1,2$. 
In the basis of number states $|n_1n_2\rangle$ the Hamiltonian decouples into two parity subsectors, $H=H_\text{even}\oplus H_\text{odd}$. Mixing can only occur by quasiparticle poisoning that will be neglected in this work\footnote{Experimentally, minimal Kitaev chains have shown poisoning times of milliseconds \cite{van_loo_single-shot_2026} with some proposals for parity qubit time predictions up to seconds \cite{karzig_quasiparticle_2021}. Quantum capacitance measurements have shown lowest readout errors for integration times of 150 $\mu s$ \cite{van_loo_single-shot_2026}.}. In the even subspace, spanned by the basis $\{|00\rangle,|11\rangle\}$ of empty or fully occupied dots, the Hamiltonian
\begin{equation}
    H_\text{even}=\begin{pmatrix}
    0&\Delta\\
    \Delta&-2\mu
\end{pmatrix}
\end{equation}
only features CAR and the average onsite energy $\mu=(\mu_1+\mu_2)/2$. The eigenstates  $H_\text{even}\ket{E_{\pm}}=\epsilon^\text{even}_{\pm}\ket{E_{\pm}}$, with  eigenenergies $\epsilon^\text{even}_\pm=-\mu\pm\epsilon/2$, are gapped by the energy difference 
\begin{equation}\epsilon=2\sqrt{\mu^2+\Delta^2}.\end{equation}
The odd subspace with one fermion in the system, which is spanned by the basis $\{|10\rangle,|01\rangle\}$, is described by the Hamiltonian
\begin{equation}
    H_\text{odd}=\begin{pmatrix}
    -\mu_1&-t\\
    -t&-\mu_2
\end{pmatrix} .
\end{equation}
Odd eigenstates $H_\text{odd}\ket{O_{\pm}}=\epsilon^\text{odd}_{\pm}\ket{O_{\pm}}$ have the energies $\epsilon^\text{odd}_{\pm}=-\mu\pm\sqrt{(\mu_2-\mu_1)^2+t^2}$. For $\mu_1=\mu_2$, the energies are  linear in $\mu$ and $t$. At the sweet spot, $\Delta=t$, both parity sectors are degenerate at $\mu=0$.

In the following, we consider a modulation of the common onsite potential, $f\to\mu$ and assume $\mu_1=\mu_2$. Fig.~\ref{fig:min_chain_susc}b shows the energy bands and quantum capacitance $C_Q$ of the MKC as a function of the common gate energy $\mu$. 
Notably, only even parity has a finite quantum capacitance of $C_Q=8(p_+-p_-)\Delta^2/\epsilon^{3}$, where $p_\pm$ is the population of eigenstate $\ket{E_\pm}$, 
which has been used to read out the parity in a gate current measurement ($A=H'$);  see 
Ref.~\cite{van_loo_single-shot_2026}.

Since we are dealing with a two-level system\footnote{Following the general definitions introduced in Sec.~\ref{sec:Linear Response in open Systems}, we use populations $p_+ = p_2$, $p_- = p_1$, energies $\epsilon_+^\text{even} = E_2$, $\epsilon_-^\text{even} = E_1$, as well as absorption $\Gamma_+=\Gamma_{21}$ and emission $\Gamma_-=\Gamma_{12}$. Dephasing $2\Gamma_\phi=\Gamma_{11}+\Gamma_{22}$ will be neglected.}, there exists a single decoherence rate $\Gamma_{T_2}=\Gamma_\phi+(\Gamma_++\Gamma_-)/2$ 
and detailed balance implies
$p_+/p_-=\Gamma_+/\Gamma_-$ for a thermalized system. 
Assuming a bosonic bath with Ohmic damping \cite{breuer_theory_2007,barr_spectral_2024},
the absorption and emission rates are
\begin{equation}
\Gamma_{\pm} 
=
|s_{\pm}|^2 \epsilon  \,\left[n_B(\epsilon)+\frac{1\mp1}{2}\right],
\end{equation} 
with the Bose-Einstein distribution $n_B(\omega)$ and the system transition matrix elements $|s_\pm|$  for absorption and emission, respectively;
see Appendix~\ref{Appendix:System-Bath}.

The even-parity and finite-frequency susceptibility in Eq.~\eqref{eq:total_susceptibility} consists of the four contributions
\begin{subequations}
\begin{align}
    \chi_\mathrm{st} &= 0\,,
    \\
    \chi_H &= \frac{8\Delta^2}{\epsilon}\frac{p_--p_+}{(\omega+i\Gamma_{T_2})^2-\epsilon^2}\,,
    \\
    \chi_\Gamma &= \frac{16\mu^2}{\epsilon^2}\frac{p_-\frac{\partial \Gamma_+}{\partial \epsilon}-p_+\frac{\partial \Gamma_-}{\partial \epsilon}}{\Gamma_-+\Gamma_+-i\omega},
    \\
    \chi_{L,\mathrm{DB}} &= -i\chi_H\frac{\Gamma_{T_2}(\omega+i\Gamma_{T_2})}{\epsilon^2}\label{eq:mkc_hermes}\, .
\end{align}
\end{subequations}
All odd-parity contributions are zero. Furthermore, $\chi_{L,\mathrm{BT}}=0$ for a thermal state since $\mathcal{G}_m = 0$.

In Fig.~\ref{fig:min_chain_susc}c we show the finite-frequency susceptibility as a function of $\mu$.
Consequently, two distinct regimes emerge, called resonant  ($\omega\approx \epsilon$) and dispersive ($\omega\ll \epsilon$). In Fig.~\ref{fig:min_chain_susc}d-e a cut at $\omega_r\approx\epsilon$ 
shows a double-peak structure mainly driven by resonant Hamiltonian contributions. Since we choose $\Gamma_{T_2}\ll\epsilon$, while Sisyphus contributions are negligible at these frequencies, 
the Hermes and Hamiltonian contributions are related via 
\begin{equation}
\chi_{L,\mathrm{DB}}(\omega_r)\simeq -i\frac{ \Gamma_{T_2}}{\epsilon} \chi_H(\omega_r).
\end{equation}
Fig.~\ref{fig:min_chain_susc}f shows the imaginary part of the susceptibility in the dispersive regime ($\omega_d\ll\epsilon$).
Both Hermes and Hamiltonian contributions are small ($\mathcal{O}(\omega/\epsilon)$) 
and for $\Gamma_{T_2}\ll\epsilon$ they are related via
$\Imag[\chi_{L,\mathrm{DB}}(\omega_d)]\simeq -\Imag[ \chi_H(\omega_d)]/2$.
At higher temperatures, the Sisyphus term becomes dominant and leads to the double-peak structure. 
In Fig~\ref{fig:min_chain_susc}g, we compare the quantum capacitance, which measures the band curvature, with the real part of the total susceptibility. 
Since
$\Real[\chi_{L,\mathrm{DB}}(\omega_d)]\simeq \Gamma_{T_2}^2 \Real[\chi_H(\omega_d)] /\epsilon^2$, the contribution
$\chi_L$ compensates for the decoherence on $\chi_H$.

\begin{figure}[t]
    \centering
    \includegraphics[width=\linewidth]{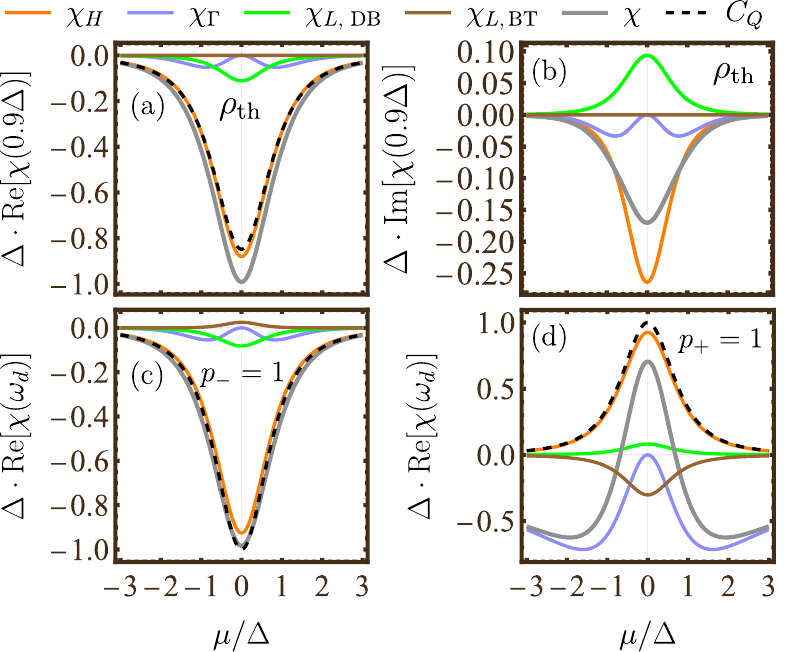}
    \caption{Susceptibility of the minimal Kitaev chain for thermal populations near resonance ($\omega\lesssim\epsilon$) for \textbf{(a)} real and \textbf{(b)} imaginary values. Real valued susceptibility when \textbf{(c)} the ground state is fully populated or \textbf{(d)} the excited state is fully populated. $\chi_{L,\mathrm{BT}}$ refers to the beyond thermalization terms of Hermes contribution while $\chi_{L,\mathrm{DB}}$ denotes detailed balance terms. Other parameters are identical to Fig.~\ref{fig:min_chain_susc}.}
    \label{fig:MKC_extra_plots}
\end{figure}

By explicitly rearranging the Hermes and Hamiltonian contributions according to Eq.~\eqref{eq:Ramon_form}, we see that 
\begin{align}
    \chi_H+\chi_{L,\mathrm{DB}} &= 8\Delta^2\frac{p_+-p_-}{\epsilon^3}\frac{\epsilon^2+\Gamma_{T_2}^2-i\omega\Gamma_{T_2}}{\epsilon^2-(\omega+i\Gamma_{T_2})^2}
    \nonumber \\
    &= C_Q+\mathcal{O}(\omega),
\end{align}
as anticipated in Eq.~\eqref{eq:dispersiveLimitHamiltonianAndHermesMainText}. We can see the effect for higher frequencies in Fig.~\ref{fig:MKC_extra_plots}a. Closer to resonance frequencies ($\omega\lesssim\epsilon$), the real part of the susceptibility overestimates the quantum capacitance. The imaginary part (Fig.~\ref{fig:MKC_extra_plots}b) grows in magnitude because it scales at least linearly with frequency. However, the effective shape of both signals remains.

When the populations do not follow a thermal distribution and detailed balance $p_+/p_- = \Gamma_+/\Gamma_-$ is broken, Hermes terms gain an additional real-valued contribution
\begin{align}
    \chi_{L,\text{BT}}&= -2i\chi_H\frac{\omega+i\Gamma_{T_2}}{\epsilon^2}\left(\,\frac{p_+\Gamma_--p_-\Gamma_+}{p_--p_+}\right),
\end{align}
which represents corrections beyond thermalization.
For either only the ground state ($p_-=1$) or excited state ($p_+=1$) populated, the sum of Hermes and Hamiltonian terms for dispersive frequencies yields
\begin{align}
    \chi_H+\chi_L=C_Q\left(1-\frac{2\Gamma_\pm\Gamma_{T_2}}{\epsilon^2+\Gamma_{T_2}^2}\right)+\mathcal{O}(\omega).
    \label{eq:beyond_thermalization_MKC_susceptibility}
\end{align}
Therefore, populations beyond thermalization will generally lead to a reduction of the pure quantum capacitance signal, as seen in Figs.~\ref{fig:MKC_extra_plots}c and \ref{fig:MKC_extra_plots}d.
When only the ground state is populated ($p_-=1$) in Fig.~\ref{fig:MKC_extra_plots}c, assuming low enough temperature, the system is still close to the thermalized state 
and $\chi_{L,\text{BT}}$ is small. However, when the excited state is populated ($p_+=1$) in Fig.~\ref{fig:MKC_extra_plots}d, $\chi_{L,\text{BT}}$ causes a significant correction and stronger disagreement between the quantum capacitance and the total susceptibility. 
Additionally, the total response is mainly masked by Sisyphus contributions that are strongly enhanced because emission rates $\Gamma_-$ are always larger than absorption rates $\Gamma_+$, resulting in strong dissipation.

This simple example already shows that corrections of populations beyond thermalization are in general not negligible, and understanding dispersive responses requires to consider the full expressions beyond curvature predictions, as presented in this work.

\subsection{Kitaev-Josephson Junction (KJJ)}
\begin{figure}[t]
    \centering
    \includegraphics[width=\linewidth]{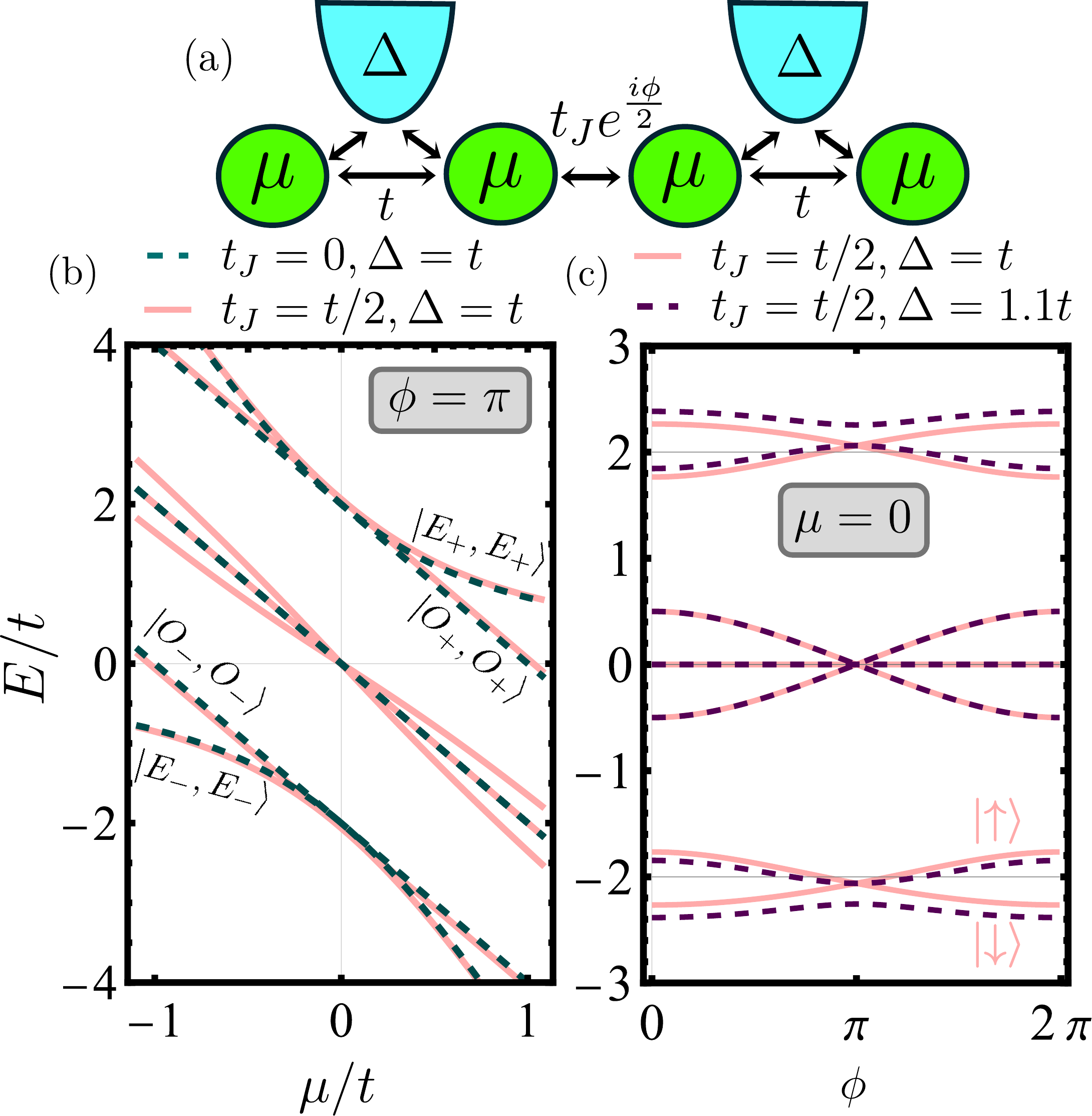}
    \caption{\textbf{(a)} Two minimal Kitaev chains tunnel coupled with $t_J$ form a Kitaev-Josephson Junction (KJJ). The superconducting leads have a phase difference of $\phi$. 
    \textbf{(b)} 
    Common gate spectrum for decoupled (green, dashed) and coupled (salmon, solid) KJJ in even global parity at $\phi=\pi$ and $\Delta=t$. 
    $|\nicefrac{E_\pm}{O_\pm},\nicefrac{E_\pm}{O_\pm}\rangle$ are many-body eigenstates of the decoupled minimal Kitaev chains corresponding to local even/odd parity. 
    At $\phi=\pi$, coupled KJJ eigenstates follow these states up to first order in $t_J$, see Eqs.~\eqref{eq:perturbation_basis}
    \textbf{(c)} Flux spectrum comparing sweet spot $\Delta=t$ (salmon, solid) with a detuned $\Delta= 1.1t$ (purple, dashed)
    system at $\mu=0$. The states the $\ket{\nicefrac{\uparrow}{\downarrow}}=\ket{E_{-},E_{-}}\pm\ket{O_{-},O_{-}}+\mathcal{O}(t_J)$ correspond to the (anti-)symmetric combinations in lowest perturbation order, see Appendix~\ref{sec:perturbative_low_regime}. When detuned from the coupling sweetspot, a gap opens.}
    \label{fig:KJJ_spectra}
\end{figure}
\begin{figure}[t]
    \centering
    \includegraphics[width=1\linewidth]{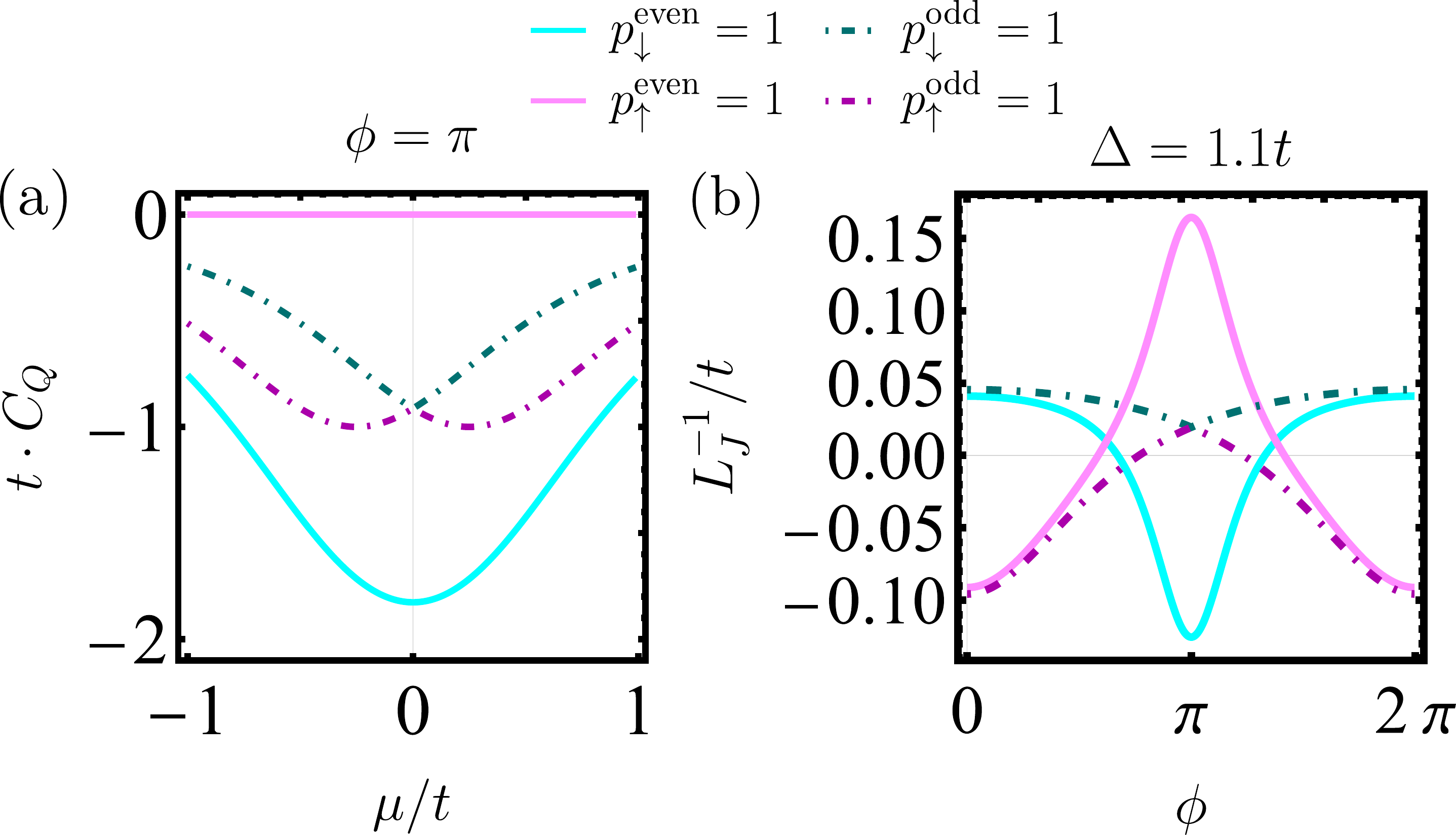}
    \caption{Curvature of the lowest two KJJ energy bands for
    \textbf{(a)} gate spectroscopy at $\phi=\pi$ and $\Delta=t$ and \textbf{(b)} flux spectroscopy at $\Delta=1.1t$ for global odd (dot-dashed lines) and even (full lines) parity. In (a) the curvature corresponds to the quantum capacitance $C_Q$, while in (b) it corresponds to the Josephson inductance $L_J$, respectively. We distinguish a full occupation of the lowest eigenstate $p_\downarrow=1$ in cyan from a full occupation of the first excited state $p_\uparrow=1$ in magenta.}
    \label{fig:KJJ_curvature}
\end{figure}
We now consider two MKCs which are coupled to each other by a phase-dependent tunnel coupling to form a Josephson junction, as sketched in Fig.~\ref{fig:KJJ_spectra}a.
First, we analyze the spectrum and energy sectors while defining a low-energy qubit space. Second, we discuss the quantum capacitance and Josephson inductance of the qubit states to finally compare them to dispersive gate and flux susceptibilities. This comparison lays the foundation for qubit readout.

\subsubsection{Spectrum and qubit definiton}
The Hamiltonian of the Kitaev-Josephson junction (KJJ) is
\begin{align}
	H_{KJJ}=H_{L}+H_{R}+H_J\,,
\end{align}
where $H_{L}$ and $H_R$ describe left and right MKCs, respectively, as introduced in Eq.~\eqref{eq:min_chain_Hamil}.
The coupling Hamiltonian
\begin{align}
	H_J = - t_J e^{i\phi/2} d_{L,2}^\dagger d_{R,1}+h.c.
\end{align}
depends on the coupling strength $t_J$ between the nearest quantum dots and the relative phase $\phi$ between the superconducting leads. 
A similar setup was first proposed in Ref.~\cite{leijnse_parity_2012} and was subsequently studied in Refs.~\cite{pino_minimal_2024,pan_rabi_2025,palacios_effect_2026}. 
The global fermionic parity of the system is conserved, 
which decouples the subspaces of global odd and even particle numbers. Given two identical MKCs with a constant onsite energy $\mu$ across the four dots, 
a weak coupling, $t_J< |t+\Delta|/2$,
allows to distinguish three  sectors in the energy spectrum: low, high and central; see Figs.~\ref{fig:KJJ_spectra}b and \ref{fig:KJJ_spectra}c for the energy spectra as a function of chemical potential $\mu$ for a fixed relative phase $\phi=\pi$ and relative phase $\phi$ for a fixed $\mu=0$, respectively. These two panels summarize both spectroscopic scenarios that are analyzed in the remainder of this work. Interestingly, at the flux sweetspot $\phi=\pi$, the KJJ can be solved exactly, see Eq.~\eqref{eq:KJJ_even_analytical} in Appendix~\ref{Appendix:KJJ_matrix}.

For global even parity, the low (high) energy sector is composed by the product state of two equal low (high)  MKC eigenstates, $\{\ket{E_{-(+)},E_{-(+)}},\ket{O_{-(+)},O_{-(+)}}\}$.
A projection onto the low-energy sector  (which we will refer to as the qubit subspace) reads
\begin{equation}
    H_Q=\begin{pmatrix}
        2 \epsilon_-^\text{even} &  t_J\frac{\Delta  }{\epsilon}\cos \left(\frac{\phi }{2}\right)\\
 t_J\frac{\Delta }{\epsilon} \cos \left(\frac{\phi }{2}\right) &  2\epsilon_-^\text{odd}
    \end{pmatrix}\,.\label{eq:qubit-space}
\end{equation}
At the flux sweet spot $\phi=\pi$, the states effectively decouple (see Appendix~\ref{sec:perturbative_low_regime} for the first-order corrections to these eigenstates originating from virtual excitations).
At the MKC sweet spot, $\mu=0$, $\Delta=t$, and $\epsilon^\text{even}_\pm=\epsilon^\text{odd}_\pm$, the eigenstates are the superpositions $\ket{\nicefrac{\uparrow}{\downarrow}}=\ket{E_{-},E_{-}}\pm\ket{O_{-},O_{-}}+\mathcal{O}(t_J)$ at energies 
$E^\text{even}_{\nicefrac{\uparrow}{\downarrow}}=\pm\frac{t_J\Delta}{\epsilon}\cos{\frac{\phi}{2}}+\mathcal{O}(t_J^2)$, which cross at $\phi=\pi$. Notably, higher-order corrections destroy the mirror symmetry of the flux bands around the MKC energy $E=\epsilon_-^\text{even}+\epsilon_-^\text{odd}$, see Fig.~\ref{fig:KJJ_spectra}c. 
Away from the MKC sweet spot, e.g., $t\neq \Delta$, the low-energy states are gapped with an anticrossing at $\phi = \pi$, as shown in Fig.~\ref{fig:KJJ_spectra}c.
The high-energy subspace is analogously described.

The global odd low- and high-energy sectors have a similar structure, but the crossing at $\phi=\pi$ remains regardless of the ratio $\Delta/t$. 
The global odd-parity spectrum is shown in Fig.~\ref{fig:KJJ_ODD} in Appendix~\ref{sec:global_odd_KJJ}. Their energies up to second order in $t_J$ can be found in Eq.~\eqref{eq:odd_low_regime_energies}. 
Notably, the spectrum in Fig.~\ref{fig:KJJ_ODD}b at the flux sweet spot $\phi = \pi$ can be solved exactly with the lowest two states following 
$E^\text{odd}_{\nicefrac{\uparrow}{\downarrow}}=-2\mu-\sqrt{\Delta^2+(\mu\pm t_J/2)^2}$, the $t_J$-shifted MKC even-parity low energy band.

The global even central energy sector features four states. Each of them is a mixture of high~$(+)$ and low~$(-)$ energy MKC eigenstates.
A projective analysis in Appendix~\ref{sec:perturbative_central_regime} reveals two degenerate states at the energy $\lambda_{1,2}=-2\mu$
and two states at $\lambda_{3,4}=-2\mu\pm t_J|\eta(\phi)|/\epsilon$ where $\eta(\phi)=\epsilon\cos(\phi/2)+2i\mu\sin(\phi/2)$;  see Figs.~\ref{fig:KJJ_spectra}b and \ref{fig:KJJ_spectra}c. 
One of the degenerate states can be identified as $\ket{D}=(\ket{O_-,O_+}-\ket{O_+,O_-})/\sqrt{2}$ featuring only local odd MKC configurations. The identity $H'_{KJJ}\ket{D}\propto\ket{D}$ qualifies it as a dark state, which is fully decoupled from the other states in the system. 

In the following, we investigate two spectroscopic scenarios  (common gate and flux) and their distinct signals in the dispersive measurement regime.
Absorption and emission are mediated by an Ohmic bath with spectral function $J(\omega)=\omega/2\pi$ leading to transition rates 
\begin{equation}
\Gamma_{mn}
=
\left\{\begin{array}{lcc}
    |s_{mn}|^2 \nu_{mn}\left(n_B(\nu_{mn})+1\right) & , & E_m>E_n\\
    |s_{mn}|^2 \nu_{nm}\,n_B(\nu_{nm}) & ,    & E_m<E_n \\
    |s_{mn}|^2 k_BT & , & E_m=E_n
\end{array}
\right. ,
\end{equation}
with $\nu_{mn}=E_m-E_n$, system-bath coupling elements $s_{mn}$ in the Hamiltonian eigenbasis, $k_B$ the Boltzmann constant, and temperature $T$. Due to parity protection within each chain, only quasiparticle poisoning can change local parity.
Poisoning times for MKCs have been shown to exceed milliseconds while dispersive measurements are performed within $\sim150 $ $\mu$s \cite{van_loo_single-shot_2026}.
Thus, it is reasonable to assume that only local parity conserving transitions occur, i.e., the left/right MKC will be locally excited by or relaxes into the bath.
The bath coupling operator [cf.~Appendix~\ref{Appendix:System-Bath}]
\begin{multline}
    S=\sum_{\substack{\pm\\P=O,E}}\Big(\ket{P_-,P_+}\bra{P_\pm,P_\pm}
    \\+\ket{P_+,P_-}\bra{P_\pm,P_\pm}+h.c.\Big)
    \label{eq:system_bath_coupling_operator}
\end{multline}
predominantly coincides with transitions between energy regimes low$\leftrightarrow$central$\leftrightarrow$high. 

\subsubsection{Curvature of the qubit states}
\label{sec:KJJ_curvatures}

First, we investigate the quantum capacitance, a measure of gate curvature; see Fig.~\ref{fig:KJJ_curvature}a.
At the flux sweet spot $\phi=\pi$, the quantum capacitance of the global even ground state 
reads
\begin{align}
    C_{Q,\downarrow}^\text{even} 
    &= -\sum_\pm\frac{\Delta^2}{\left(\Delta ^2+\left(\mu\pm \frac{t_J}{2}\right) ^2\right)^{3/2}}
    \nonumber \\
    &\approx
    -\frac{16\Delta^2}{\epsilon^3}+96\,t_J^2\Delta^2\left(\frac{\Delta^2-4\mu^2}{\epsilon^7}\right),
\end{align}
(see all eigenenergies in Appendix~\ref{Appendix:KJJ_matrix})
corresponding to the sum of two MKC ground state curvatures shifted by $\pm t_J/2$. The first excited state has no curvature, $C_{Q,\uparrow}^\text{even}=0$. Effectively, the flux sweet spot turns off the interaction between both MKCs.
For the global odd sector,
the two lowest-energy states cross at $\mu=0$. Their quantum capacitances read 
\begin{align}
    C_{Q,\nicefrac{\uparrow}{\downarrow}}^\text{odd}
    &= 
    -\frac{\Delta^2}{\left(\Delta ^2+\left(\mu\pm \frac{t_J}{2}\right) ^2\right)^{3/2}}
    \nonumber \\
    &\approx
    -\frac{8\Delta ^2}{\epsilon^3}\pm\frac{48 \Delta ^2 \mu  t_J}{\epsilon^5}+\frac{48 \Delta ^2
   t_J^2 \left(\Delta ^2-4 \mu ^2\right)}{\epsilon^7},
\end{align}
corresponding to a single MKC quantum capacitance signal shifted by $\pm t_J/2$. 
In conclusion, these quantum capacitance measurements would predict both the global and local parity while operated at the sweet spot.

We now investigate the inverse Josephson inductance, a measure of flux curvature, for fixed $\mu=0$. We present approximate analytical solutions for the lowest two eigenstates in Appendix \ref{sec:perturbative_low_regime}; Eq.~\eqref{eq:global_even_flux_curvature} for global even and Eq.~\eqref{eq:global_odd_flux_curvature} global odd parity.
The exact inverse Josephson inductances are depicted in Fig.~\ref{fig:KJJ_curvature}b. As global even and global odd parity are degenerate for $t=\Delta$, also their curvatures are. As expected, the Josephson induction of the ground and excited state are not symmetric to each other because of corrections from high-energy states, see Appendix~\ref{sec:perturbative_low_regime}. When $t\neq\Delta$, the global odd case only modulates slightly, while the global even parity shows opposite peaks emerging around $\phi=\pi$. This allows us to distinguish between global odd and global even parity as well as between global even parity eigenstates.

In the following, both the quantum capacitance and the inverse Josephson induction will provide a comparison for dispersive frequency common gate and flux susceptibilities, respectively. We will focus on the global even parity sector, because these two lowest energy states could be distinguished via both spectroscopic scenarios.

\subsubsection{Common gate response}

As a natural continuation of minimal Kitaev chains, we simulate common gate spectroscopy on the global even parity KJJ. Therefore, the modulated parameter $f$ is the chemical potential, $f = \mu$, and the read-out observable is the common gate current $A=H'_{KJJ}$. Since $A(\mu)=A$, $\chi_{\text{st}}=0$.

\begin{figure}[t]
    \centering
    \includegraphics[width=1\linewidth]{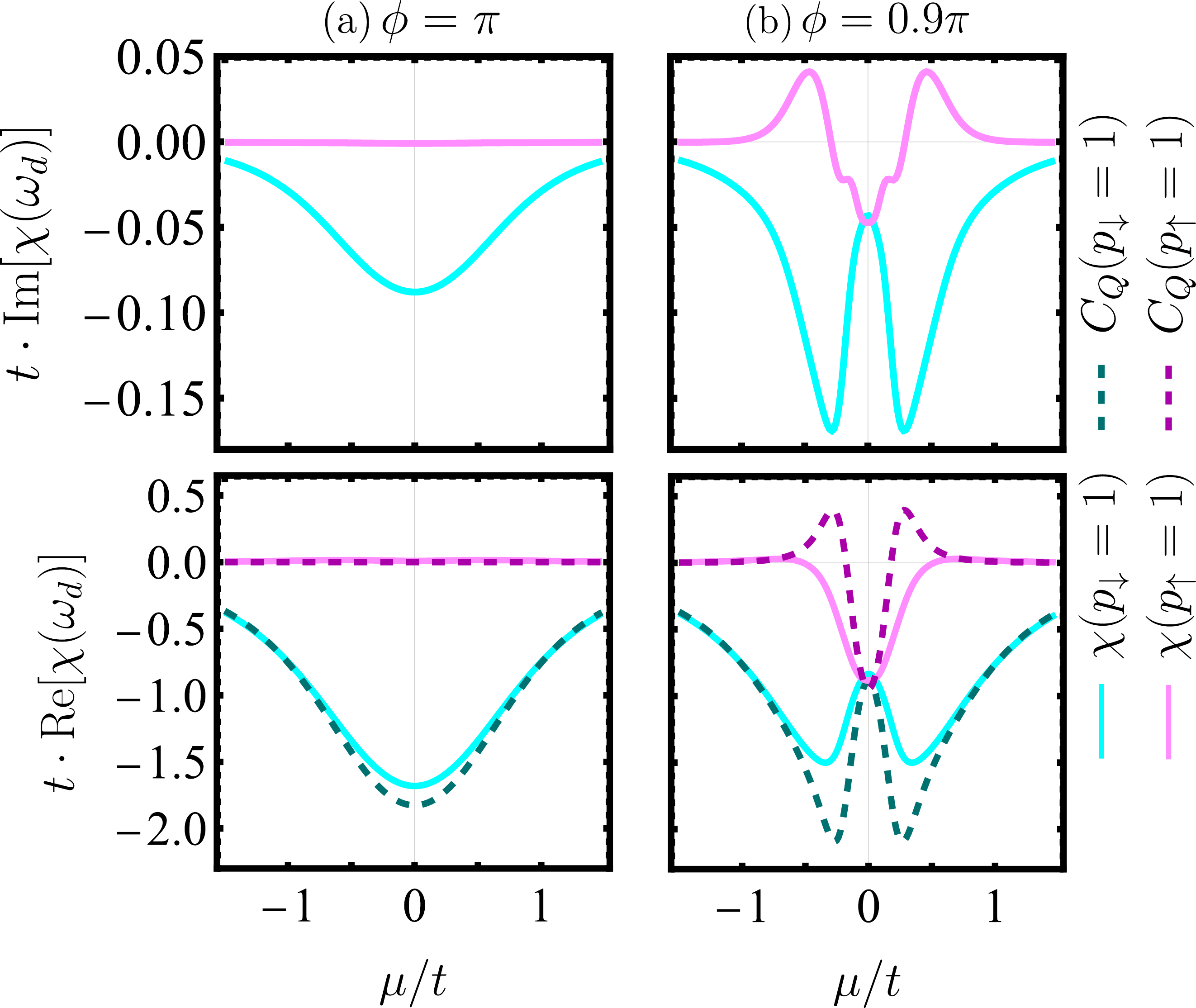}
    \caption{ Common gate susceptibility for global even parity KJJ at dispersive frequency $\omega_d$ \textbf{(a)} at the sweetspot $\phi=\pi$ and \textbf{(b)} close to the sweetspot $\phi=0.9\pi$. We depict the occupation of the ground state $p_\downarrow=1$ in green and the first excited state $p_\uparrow=1$ in orange. In the real part, we compare quantum capacitance in dashed lines. Parameters: $t_J=t/2, \Delta=t, \omega_d=t/5, k_BT=4\Delta/5$.}
    \label{fig:KJJ_gate}
\end{figure}

In Fig.~\ref{fig:KJJ_gate}, we compare dispersive measurements at $\omega = t/5$ in the global even parity sector at flux $\phi = \pi$ (sweet spot) with measurements at $\phi \neq \pi$ away from the sweet spot.
At $\phi=\pi$ we see that the susceptibilities follow the behavior of the quantum capacitance, as shown in Fig.~\ref{fig:KJJ_gate}a and consistent with Eq.~\eqref{eq:dispersiveLimitHamiltonianAndHermesMainText}. Since interactions within the qubit space are swiched off, the main response originates from local parity conserving transitions between adjacent energy regimes.
The ground state ($p_\downarrow=1$) shows a slightly lower magnitude than the quantum capacitance $C_Q$ would predict because non-thermalized populations cause additional Hermes contributions $\chi_{L,\text{BT}}$ that are not suppressed by dispersive frequencies. This can be seen in detail in Appendix~\ref{Appendix:Susceptibility_contributions_KJJ_gate} where we show the total susceptibility separated into its contributions.
The signal of the first excited state ($p_\uparrow=1$) is negligible. 

Interestingly, when a band gap between the lowest energy states opens by detuning slightly away from $\phi=\pi$ (Fig.~\ref{fig:KJJ_gate}b), a qubit signal emerges as a double-peak structure in Hamiltonian and Hermes contributions. 
The quantum capacitance overestimates the magnitude of these peaks. The Hermes terms beyond thermalization become more prominent because $|(\mathcal{G}_m-\mathcal{G}_n)/(E_m-E_n)|\approx1$ for small gaps and thus affect the real part of the susceptibility. At $\mu=0$, both populations become indistinguishable. Therefore, a qubit population measurement can only be performed when $\mu\neq0$.

\subsubsection{Flux response}

\begin{figure}
    \centering
    \includegraphics[width=\linewidth]{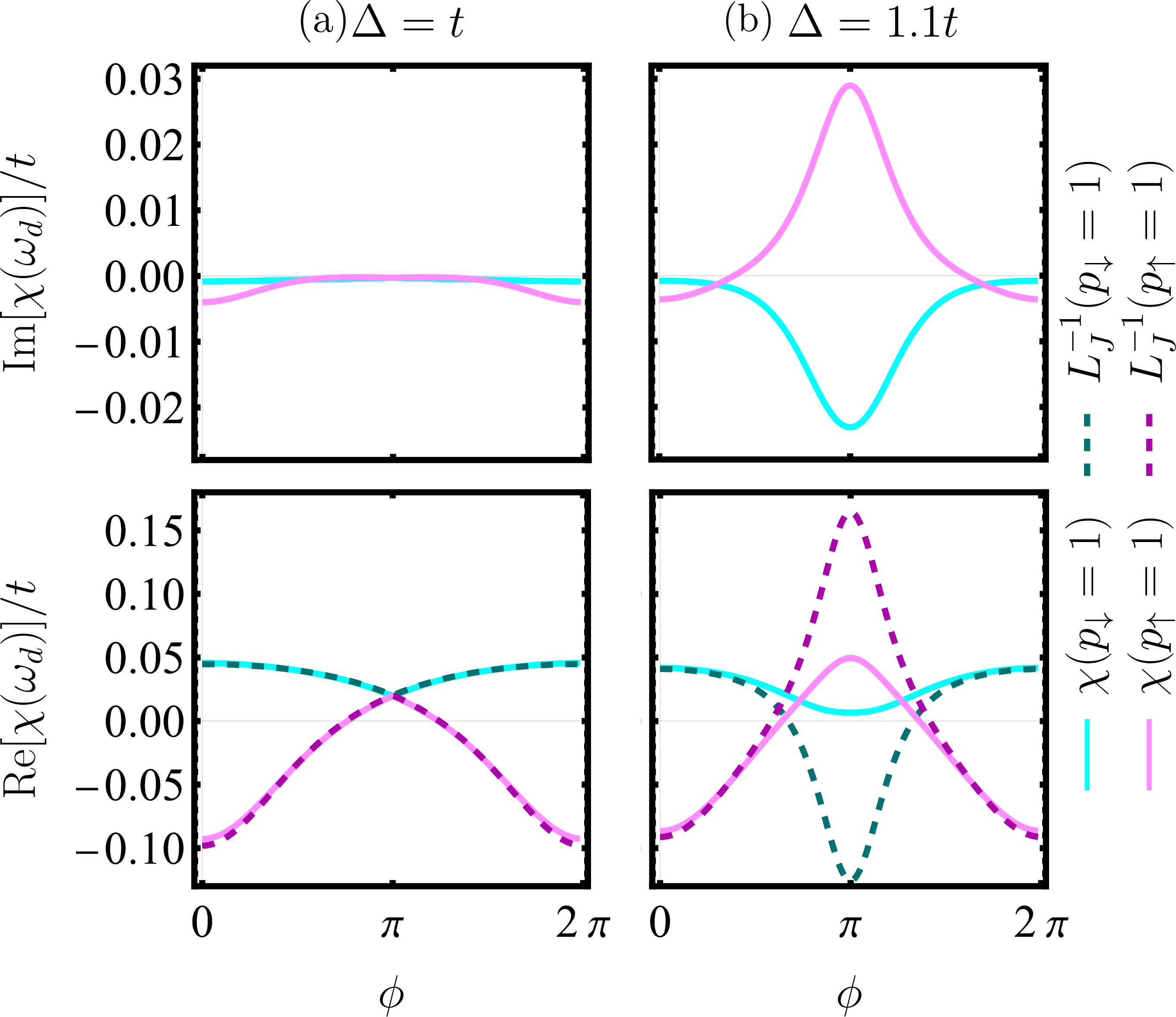}
    \caption{Flux susceptibility for ground ($p_\downarrow$, cyan) and excited ($p_\uparrow=1$, magenta) states at dispersive frequency $\omega_d$ \textbf{(a)} in coupling sweetspot $\Delta=t$ and away from the coupling sweetspot $\Delta=1.1t$. We also compare with inverse Josephson inductance in dashed lines for the real part. 
    Parameters: $k_BT=4\Delta/5,t_J=t/2,\omega_d=t/5,\mu=0$. }
    \label{fig:KJJ_flux}
\end{figure}

Introducing flux modulations $f\to\phi$ and reading out the supercurrent along the junction $A=I_S\propto H'_{KJJ}$ will yield the flux susceptibility across the KJJ shown in Fig.~\ref{fig:KJJ_flux}. 
In contrast to gate spectroscopy, the static contribution $\chi_\text{st}$ (Eq.~\eqref{eq:static_susceptibility}) leads to a frequency-independent offset to the real part of the susceptibility since $A(\phi)$ is flux-dependent.
This is known as diamagnetic current in Refs.~\cite{trif_dynamic_2018,dassonneville_dissipation_2013}.

When couplings are aligned, $\Delta=t$ in Fig.~\ref{fig:KJJ_flux}a, the imaginary part of the susceptibility is negligible around the sweet spot
while the real part follows the inverse Josephson induction of the lowest two energy bands. Importantly, at the flux sweet spot $\phi=\pi$, the susceptibility of both states is identical because the flux spectrum is degenerate. 

Detuning the couplings, $\Delta\neq t$, avoids the low-energy band crossing, as shown in Fig.~\ref{fig:KJJ_spectra}c. Hence, in Fig.~\ref{fig:KJJ_flux}b, peaks in the susceptibility emerge which allow to distinguish the parity of the system.
While a fully populated ground state leads to a dip, a fully populated first excited state results in a peak. These peaks are distinct in the imaginary part of susceptibility.
In the real part, the inverse Josephson induction predicts large peaks at $\phi=\pi$ as well. However, the susceptibility, in analogy to the result Eq.~\eqref{eq:beyond_thermalization_MKC_susceptibility} of a single MKC, is reduced. 
While Eq.~\eqref{eq:dispersiveLimitHamiltonianAndHermesMainText} predicts the static, Hamiltonian, and Hermes terms to approach the curvature at dispersive frequencies, Hermes contributions beyond thermalization $\chi_{L,\mathrm{BT}}$ compensate the peak. This keeps the signal small. A more detailed analysis of each contribution is presented in Appendix~\ref{Appendix:Susceptibility_contributions_KJJ_flux}.

The global odd response does not change significantly when detuning $\Delta\neq t$, as discussed in Sec.~\ref{sec:KJJ_curvatures}, and therefore looks similar to Fig.~\ref{fig:KJJ_flux}a.\\

To distinguish global parity, we can thus measure the susceptibility at $\phi=\pi$. In the imaginary part, the lowest two global odd eigenstate responses nearly vanish. The lowest two global even eigenstate responses yield opposite peaks. Similarly, detection can be done in the real part, however, in our case, beyond thermalization effects significantly suppress the peaks of the global even response.

\subsubsection{Comparison of qubit readout strategies}
We have now introduced common gate and flux susceptibility measurements, which allow us to distinguish between different qubit occupations. 

Gate spectroscopy allows for the distinction of global odd from global even parity. Within global even parity, we can discern local even and local odd parity as well. 
Spectroscopy can be performed even when the KJJ is tuned to the sweet spot because MKC transitions are dominant, while qubit transitions are suppressed. However, common gate readout requires a simultaneous drive and control of local onsite energies. The common drive of minimal Kitaev chains was achieved in Ref.~\cite{van_loo_single-shot_2026} by driving the common superconducting lead, which shifts Andreev bound states in the proximitized gap.
For the KJJ, both superconducting leads would need to be driven simultaneously. Interestingly, local parity readout on only one of the chains does not produce distinguishable signals at the sweet spot; see Appendix~\ref{Appendix:Left_chain}. 

In contrast, flux is a global variable controlling the phase difference between the superconducting leads of a Josephson junction. Shunting a transmon circuit with a KJJ creates the Kitmon qubit, thus flux control is already integrated into the architecture. A flux drive can be measured through the supercurrent of superconducting leads which reveals flux susceptibilities. Distinct susceptibilities close to the flux sweet spot only arise if $\Delta\neq t$.

Differences between individual parity measurements are especially prominent in the imaginary part of susceptibility. A global odd parity results in a vanishing susceptibility while, for a global even parity ground state, a negative peak and, for the first excited state, a positive peak emerges. 
However, detuning the system away from the sweet spot $t = \Delta$ might cause decreased coherence times, since a gap is opened only in the global even parity sector and local parities mix due to unprotected transitions.

\section{Conclusions}
\label{sec:Conclusions}

In this work, we have established a general framework of a Lindblad-based linear response theory for a general number of energy levels $N$, any parameter modulation $f$, observable $A$, and reference state $\rho$.
In this context, we derived the finite-frequency susceptibility where contributions of Sisyphus and Hermes terms exhibit a more nuanced dependence on the dissipator and the populations when compared with previous results \cite{peri_unified_2024}.

The central conceptual insight is that the Hermes susceptibility suppresses the effect of pure dephasing and even dissipation for thermalized occupations in the dispersive regime, enforcing that the total susceptibility follows the curvature of the energy bands, see Eq.~\eqref{eq:dispersiveLimitHamiltonianAndHermesMainText}. Additionally, we formally identify deviations of the susceptibility from curvature predictions for non-thermalized populations and introduce $|(\mathcal{G}_m-\mathcal{G}_n)/(E_m-E_n)|\ll1$ with measure of thermalization $\mathcal{G}_m$ as an important threshold for when curvature predictions suffice.

Applied to minimal Kitaev chains, these ideas are verified in Fig.~\ref{fig:min_chain_susc}g as Hermes contributions ensure the real part of the susceptibility 
recovers the band curvature even with dissipation when populations are thermalized. Going beyond thermalized populations breaks this recovery and suppresses the signal, as illustrated in Fig.~\ref{fig:MKC_extra_plots}d.

Coupling two minimal Kitaev chains creates a Kitaev-Josephson junction, which provides the superconducting phase difference as another control parameter beyond onsite energies. As a result, both common gate and flux dispersive measurements may be used to distinguish global and local parity.
In common gate spectroscopy mainly local parity preserving transitions, inherited from the individual minimal Kitaev chains, dominate the response, as demonstrated in Fig.~\ref{fig:KJJ_gate}a. While flux spectroscopy probes the qubit space directly, qubit occupation are only distinguishable for $\Delta\neq t$, see Fig.~\ref{fig:KJJ_flux}b. However, the dispersive signals are suppressed for non-thermalized states. Hence, full knowledge of the finite-frequency response beyond curvature-based predictions become indispensable to identify each qubit state.

These conclusions extend beyond the Kitaev systems studied here and apply generally to qubit architectures employing curvature-based dispersive readout \cite{van_loo_single-shot_2026,caceres_ferbo_2026,pino_minimal_2024,dourado_assessing_2026,jennings_probing_2025}. Whenever the system is not thermalized, the full finite-frequency response must be considered for a reliable interpretation of dispersive measurements.


\section{Acknowledgements}
\noindent Thanks to my friends and family for moral support.
We acknowledge funding from the
Horizon Europe Framework Program of the European
Commission through the European Innovation Council
Pathfinder Grant No. 101115315 (QuKiT). R.A. acknowledges the Spanish Ministry of Science, Innovation, and Universities through Grants CEX2024-001445-S (Severo Ochoa Centres of Excellence program) and PID2024-161156NB-I00.
We acknowledge useful discussions with M. Freudig, M. J. Calder\'on, L. Peri and F. Gonzalez-Zalba.

	\appendix
	
\onecolumngrid

\section{Derivation of Lindblad-based linear response theory}
\label{AppendixA}

\subsection{The Lindblad equation of a system coupled to a bosonic bath}
\label{Appendix:System-Bath}
In order to arrive at an effective Lindblad description of the time evolution of the reduced density matrix $\rho_S$ of an open quantum system, we use the Caldeira-Leggett model \cite{caldeira_quantum_1983} and start with a general Hamiltonian of the form $H = H_S + H_B + V$.
We assume that the system has $N$ energy levels and that the Hamiltonian takes the diagonal form $H_S = \sum_{m=1}^N E_m \ket{\phi_m}\bra{\phi_m}$, with energies $E_m$ and eigenstates $\ket{\phi_m}$, while $H_B = \sum_{k=0}^\infty \omega_k b^\dag_k b_k$ describes a bosonic bath of infinitely many independent harmonic oscillators with energies $\omega_k$ and annihilation (creation) operators $b^{(\dag)}_k$.
The coupling between the system and the bath is assumed to take the form $V = S \otimes B$, where $S = \sum_{m,n=1}^N s_{mn} \ket{\phi_m}\bra{\phi_n}$ and $B = \sum_{k=0}^\infty g_k (b^\dag_k + b_k)$ with system transition matrix elements $s_{mn}$ and system-bath couplings $g_k$.

We follow the standard derivation for open quantum systems in which we trace out the bath degrees of freedom within a Born-Markov approximation \cite{breuer_theory_2007}, i.e., the system-bath coupling is assumed to be weak and memory effects are negligible.
By a secular approximation that neglects rapidly oscillating terms and by neglecting the so-called Lamb-shift that renormalizes the energy spectrum of the system, we arrive at an effective Lindblad equation $\partial_t \rho_S = \mathcal{L}\rho_S$, where the Lindblad superoperator $\mathcal{L}$ is time-independent and given by 
\begin{align}
    \label{eq:LindbladWithDissipator}
    \mathcal{L}\rho_S = - i [H_S , \rho_S] + \mathcal{D}\rho_S,
    \qquad 
    \mathcal{D}\rho_S=\sum_{m,n = 1}^N \Gamma_{mn}\left(L_{mn}\rho_SL_{mn}^\dagger-\frac{1}{2}\big\{L^\dagger_{mn}L_{mn},\rho_S\big\}\right) ,
\end{align}
which is also presented in Eq.~\eqref{eq:LindbladianGeneralMainText} in the main text.
The dissipation superoperator $\mathcal{D}$ takes into account the interaction with the bosonic bath, which generally leads to decoherence in the system. 
Furthermore, $L_{mn} = \ket{\phi_m}\bra{\phi_n}$ are general jump operators that describe transitions from state $\ket{\phi_n}$ to state $\ket{\phi_m}$ with environmental rates 
\begin{equation}
\Gamma_{mn} = |s_{mn}|^2 \, C(E_n - E_m)\,.\end{equation}
Here, we introduced the Fourier transform of the bath correlation function
\begin{align}
    \label{eq:bathCorrelationFunctionAppendix}
    C(\omega) =  
    \int_{-\infty}^\infty \braket{B(t) B(0)}_B \, e^{i\omega t} dt
    =
    \left\{
    \begin{array}{lll}
        2\pi\, J(\omega) \, [1 + n_B(\omega) ] &,& \omega > 0
        \\
        2\pi\,  J(|\omega|) \, n_B(|\omega|) & , & \omega < 0
    \end{array} 
    \right. ,
\end{align}
where $J(\omega) = \sum_{k=0}^\infty g_k^2 \delta(\omega-\omega_k)$ and $n_B(\omega) = (e^{\beta \omega} - 1)^{-1}$ for $\omega > 0$ are the 
spectral function and bosonic thermal population of the bath modes, respectively, with the inverse temperature $\beta = 1/(k_B T)$ and the Boltzmann constant $k_B$. 

In this work, we assume that the bath is Ohmic, meaning that $J(\omega) \propto \omega$ at small energies \cite{barr_spectral_2024,breuer_theory_2007}, resulting in the well-defined finite limit $C(0) < \infty$.
Therefore, the limit $E_n = E_m$ represents pure dephasing of the energy level $E_m$ with finite rates $\Gamma_{mm} \propto |s_{mm}|^2$. 
In addition, $\Gamma_{mn}$ for $E_n > E_m$ describes emission of energy into the bath, while the case $E_n < E_m$ describes absorption of energy from the bath, which is also expressed in the detailed balance condition $\Gamma_{nm}/\Gamma_{mn} = e^{-\beta(E_n - E_m)}$.
Note that in the zero-temperature limit $\beta \to \infty$ with $n_B \to 0$, the system only emits energy into the bath and all absorption rates are zero.

As a remark, note that dephasing is also sometimes introduced as pairwise Pauli-$z$-like operators $\sigma_z^{mn} \propto L_{mm} - L_{nn}$ to draw a connection to the dephasing master equation commonly used for two-level systems in quantum optics and quantum information \cite{peri_unified_2024,navarrete-benlloch_introduction_2022}.
As this over-complicates the notation and only introduces a lot of over-counting of dephasing contributions from each energy level, resulting in a trivial redefinition of the dephasing contribution to the decoherence rates [cf.~Eq.~\eqref{eq:general_Gamma_T_2} below], we introduce dephasing with a single operator $L_{mm}$ for each energy level, leading to a more natural, unified, and clean notation.

In the rest of the appendix, as well as in the main text, we drop the label $S$ for both the density matrix and the Hamiltonian of the system to simplify the notation.

\subsection{Linear response}
\label{Appendix:expectation_value}
In this section, we provide an instructive derivation of the Lindblad-based Kubo formula as it was used in \cite{albert_geometry_2016, ban_linear_2015, wei_linear_2011,campos_venuti_dynamical_2016,villegas-martinez_application_2016}. 
In general, the expectation value $\langle A \rangle(t) = \mathrm{tr}\{ A \rho(t) \}$ of an operator $A$ is calculated from the solution $\rho(t)$ of the Lindblad equation $\partial_t \rho(t) = \mathcal{L}\rho(t)$ that was derived in the previous section. 
In the following, we assume that we perturb a given parameter $f$ of the system by a time-periodic small modulation $\delta f(t)$, such that $f \to f + \delta f(t)$.
Consequently, by assuming that all quantities depend on this parameter $f$, this generates time-dependent first-order corrections to the density matrix, $\rho(t) \to \rho(t) + \delta\rho(t)$, the Lindblad superoperator $\mathcal{L} \to \mathcal{L} + \mathcal{L}' \, \delta f(t)$, and the operator $A \to A + A' \delta f(t)$, where prime denotes the derivative with respect to said parameter $f$.
In this sense, the full expectation value of the operator $A$ to first order becomes $\langle A \rangle(t) \to 
    \langle A \rangle(t) + \delta\langle A \rangle(t)$, with the first-order correction
\begin{align}
    \delta\langle A \rangle(t)
    =
    \delta f(t) \, \mathrm{tr}\{ A' \rho(t) \} +\mathrm{tr}\{ A \, \delta\rho(t) \} ,
\end{align}
for which we need an equation that determines the change in the density matrix $\delta\rho(t)$.
This change follows from the Lindblad equation $\partial_t \rho(t) = \mathcal{L}\rho(t)$, from which we obtain the first-order correction 
\begin{align}
    \partial_t \, \delta \rho(t) 
    -
    \mathcal{L}  \, \delta\rho(t)
    = 
    \delta f(t) \, \mathcal{L}'\rho(t),
\end{align}
which is an inhomogeneous, first-order differential equation for the correction $\delta \rho(t)$. 
Using standard integration techniques and assuming $\delta \rho(-\infty) = 0$ as initial condition, the solution takes the form
\begin{align}
    \delta \rho(t)
    =
    \int_{-\infty}^t  e^{(t-\tau)\mathcal{L}}\mathcal{L}'\rho(\tau) \, \delta f(\tau) \, d\tau ,
\end{align}
with which we arrive at
\begin{align}
\label{eq:appendixLinearResponseAchiF}
    \delta\langle A \rangle(t)
    =
    \int_{-\infty}^\infty 
    \chi(t,\tau) \, \delta f(\tau) \, d\tau,
\end{align}
where 
\begin{align}
\chi(t,\tau) = 
    \delta(t-\tau) \, \mathrm{tr}\{ A' \rho(\tau) \}  
    +  
    \Theta(t-\tau) \, \mathrm{tr}\{ A \, e^{(t-\tau)\mathcal{L}} \mathcal{L}'\rho(\tau)   \} 
\end{align}
is the linear-response susceptibility. 

For arbitrary time-dependent states $\rho(t)$, $\chi(t,\tau)$ is nonlocal in time and depends on $t$ and $\tau$ separately. 
By assuming that $\rho(t) \equiv \rho$ is approximately constant over the time scale of the linear-response measurement, the susceptibility becomes time-local, $\chi(t,\tau) = \chi( t - \tau)$, and simplifies to
\begin{align}
\chi(t) = 
    \delta(t) \, \mathrm{tr}\{ A' \rho \}  
    +  
    \Theta(t) \, \mathrm{tr}\{ A \, e^{t\mathcal{L}}  \mathcal{L}' \rho    \} ,
    \label{eq:totalLinearResponseFunctionAppendix}
\end{align}
with an arbitrary, initially prepared state described by the density matrix $\rho$.
This is the result presented in Eqs.~\eqref{eq:ulr_Kubo-formula} and \eqref{eq:ulr_susceptibility_general} in the main text.

As a last step, we specify the terms appearing in the contribution due to the change of the Lindblad superoperator $\mathcal{L}'$. Following Eq.~\eqref{eq:LindbladWithDissipator}, such changes generally originate from changes in the system Hamiltonian $H'$, the transition rates $\Gamma_{mn}'$, or the jump operators $L_{mn}'$. 
In other words, we can split $\mathcal{L}' = \mathcal{L}_H' + \mathcal{L}_\Gamma' + \mathcal{L}_L'$ with
\begin{subequations}
\label{eq:SusceptContributionAppendix}
\begin{align}
    \label{eq:SusceptContributionAppendixHamiltonian}
    \mathcal{L}_H'\rho &= - i \left[ H' , \rho \right] ,
    \\
    \label{eq:SusceptContributionAppendixSisyphus}
    \mathcal{L}_\Gamma'\rho
    &= 
    \sum_{m,n = 1}^N \Gamma_{mn}' \left(L_{mn} \rho L_{mn}^\dag - \frac{1}{2}\big\{L^\dag_{mn}L_{mn},\rho \big\}\right) ,
    \\
    \label{eq:SusceptContributionAppendixHermes}
    \mathcal{L}_L'\rho
    &=
    \sum_{m,n = 1}^N \Gamma_{mn} 
    \left( L_{mn}' \rho L_{mn}^\dag
    +
    L_{mn}\rho (L_{mn}^\dag)'
    -
    \frac{1}{2}\left\{
    (L_{mn}^\dag)' L_{mn}
    +
    L_{mn}^\dag
    L_{mn}'
    ,\rho 
    \right\}
    \right),
\end{align}
\end{subequations}
resulting in the decomposition of the susceptibility into static, Hamiltonian, Sisyphus, and Hermes contributions, as presented in Eqs.~\eqref{eq:total_susceptibility} to  \eqref{eq:other_contributions_susceptibility} in the main text.

\subsection{Diagonalization of the Lindblad superoperator}
\label{section:diagonalizationSuperoperatorAppendix}
In order to efficiently calculate the total susceptibility $\chi(t)$ in Eq.~\eqref{eq:totalLinearResponseFunctionAppendix}, we need to apply the superoperator $e^{t\mathcal{L}}$ to the operator $\mathcal{L}'\rho$. 
For this, we diagonalize the nonhermitian Lindblad superoperator $\mathcal{L}$ to find its eigenvalues $\lambda \in \mathbb{C}$, such that $\mathcal{L} r = \lambda r$ and $\mathcal{L}^\dag l = \lambda^* l$ with right and left eigenoperators $r$ and $l$, respectively. 
The adjoint superoperator $\mathcal{L}^\dag$ is defined via $\braket{\mathcal{L}^\dag X,Y} = \braket{X,\mathcal{L}Y}$ for system operators $X$ and $Y$, where $\braket{X,Y} = \mathrm{tr}(X^\dag Y)$ is the Hilbert-Schmidt inner product.
A short straightforward calculation using cyclic permutation under the trace shows that if 
\begin{align}
    \mathcal{L}Y = - i [H , Y] + 
    \sum_{m,n = 1}^N \Gamma_{mn}\left(L_{mn} Y L_{mn}^\dagger-\frac{1}{2}\big\{L^\dagger_{mn}L_{mn},Y\big\}\right) ,
\end{align}
as given by Eq.~\eqref{eq:LindbladWithDissipator}, then
\begin{align}
    \mathcal{L}^\dag X =  i [H,X] + \sum_{m,n = 1}^N \Gamma_{mn}\left(L_{mn}^\dag X L_{mn} - \frac{1}{2}\big\{L^\dagger_{mn}L_{mn},X\big\}\right) .
\end{align}

An operator $X$ of the system described by the Hamiltonian $H$ can be written in terms of the general jump operators $L_{mn} = \ket{\phi_m}\bra{\phi_n}$ as $X = \sum_{m,n=1}^N x_{mn} L_{mn}$ with coefficients $x_{mn} = \braket{\phi_m|X|\phi_n}$ in terms of eigenstates $\ket{\phi_m}$ of $H$. 
Therefore, it is enough to know how both $\mathcal{L}$ and $\mathcal{L}^\dag$ act on the basis operators $L_{mn}$.
We find
\begin{align}
    \label{eq:appendix_howLindbladActsOnLmn}
    \mathcal{L} L_{mn} 
    = 
    \lambda_{mn}  L_{mn} 
    +
    \delta_{mn} 
    \sum_{k=1}^N \Gamma_{km} L_{kk} 
    ,
    \qquad 
    \mathcal{L}^\dag L_{mn} 
    = 
    \lambda_{mn}^*  L_{mn} 
    +
    \delta_{mn} 
    \sum_{k=1}^N \Gamma_{mk} L_{kk} 
    ,
\end{align}
with $\lambda_{mn} = - \Gamma_{T_2}^{mn} - i (E_m-E_n)$ and the decoherence rates 
\begin{align}
    \label{eq:general_Gamma_T_2}
    \Gamma_{T_2}^{mn} = \frac{1}{2}\sum_{k=1}^N (\Gamma_{km} +\Gamma_{kn}).
\end{align}
For all $m$ and $n$, both $\mathcal{L}$ and $\mathcal{L}^\dag$ can generally be represented as $N^2 \times N^2$ matrices.
However, as we discuss below, there is only need to diagonalize the $N\times N$ sub-block for $m=n$, since the superoperator is already diagonal for $m \neq n$.

\subsubsection{Case $m \neq n$}
\label{section:diagonalizationSuperoperatorAppendixMnotN}
For $m \neq n$, we immediately see from Eq.~\eqref{eq:appendix_howLindbladActsOnLmn} that the jump operators $L_{mn}$ are the left and right eigenoperators, $l_{mn} = r_{mn} = L_{mn}$, with the eigenvalues $\lambda_{mn}$, and they are orthonormal with respect to the Hilbert-Schmidt inner product, $\braket{L_{kl},L_{mn}} = \delta_{km} \delta_{ln}$.

\subsubsection{Case $m = n$}
\label{section:diagonalizationSuperoperatorAppendixMequalN}
For $m = n$, Eq.~\eqref{eq:appendix_howLindbladActsOnLmn} reduces to the two systems of $N$ coupled equations
\begin{align}
    \label{eq:leftRightEquationsLindblad}
    \mathcal{L} L_{mm} 
    = 
    - \sum_{k=1}^N \Gamma_{km} L_{mm} 
    + 
    \sum_{k=1}^N \Gamma_{km} L_{kk} 
    ,
    \qquad 
    \mathcal{L}^\dag L_{mm} 
    = 
    - \sum_{k=1}^N \Gamma_{km}  L_{mm} 
    + 
    \sum_{k=1}^N \Gamma_{mk} L_{kk} 
    .
\end{align}
We see that these equations are not yet diagonal, and we need to solve the eigenvalue equations $\mathcal{L}r_m = \lambda_m r_m$ and $\mathcal{L}^\dag l_m = \lambda_m l_m$, where $r_m$ and $l_m$ are the right and left eigenoperators satisfying $\braket{l_m,r_n} = \delta_{mn}$.
The corresponding eigenvalues $\lambda_m \leq 0$ will be strictly real and non-positive. A unique zero eigenvalue corresponds to a thermal state for which the right eigenoperator is the thermal equilibrium density matrix and the left eigenoperator is the identity.

In order to find the eigenoperators and eigenvalues numerically for an arbitrary number of system states $N$, we construct the $N \times N$ dissipation matrices $(\mathcal{L}_{ab})$ and  $(\mathcal{L}_{ab}^\dag)$  with matrix elements
\begin{align}
    \mathcal{L}_{ab} 
    = 
    \braket{L_{aa},\mathcal{L}L_{bb}} 
    = 
    \Gamma_{ab}  
    - \delta_{ab} \sum_{k=1}^N \Gamma_{kb}   
     ,
    \qquad 
    \mathcal{L}^\dag_{ab} =
    \braket{L_{aa},\mathcal{L}^\dag L_{bb}}
    =   
    \Gamma_{ba}-  \delta_{ab} \sum_{k=1}^N \Gamma_{kb},
\end{align}
in which dephasing rates $\Gamma_{aa}$ are absent and only dissipative absorption and emission rates $\Gamma_{ab}$ for $a\neq b$ enter.
Note that the dissipation matrices also satisfy $(\mathcal{L}_{ab})^\mathrm{T} = (\mathcal{L}_{ab}^\dag)$.

\subsubsection{Example: Diagonalization of the dissipation matrices for a  two-level system}
\label{sec:appendixExampleTLS}
For a two-level system with $N=2$ states, the dissipation matrices in the basis $\{L_{11},L_{22}\}$ are
\begin{align}
    (\mathcal{L}_{ab}) 
    =
    \begin{pmatrix}
        - \Gamma_{21} & \Gamma_{12}
        \\
        \Gamma_{21} & -\Gamma_{12}
    \end{pmatrix}
    =
    (\mathcal{L}_{ab}^\dag)^\mathrm{T}. 
\end{align}
The eigenvalues are $\lambda_{1} = 0$ and $\lambda_2 = - \Gamma_{12} - \Gamma_{21}$. 
For the eigenvalue $\lambda_1 = 0$, the left and right eigenvectors are
\begin{align}
    l_1 \propto \begin{pmatrix}
        1 \\ 1
    \end{pmatrix},
    \qquad 
    r_1 \propto \begin{pmatrix}
        \Gamma_{12} \\ \Gamma_{21}
    \end{pmatrix},
\end{align}
corresponding to the normalized eigenoperators
\begin{align}
    l_1 = L_{11} + L_{22} \equiv \mathbbm{1},
    \qquad
    r_1 = \frac{ \Gamma_{12} }{ \Gamma_{12} + \Gamma_{21} } L_{11} + \frac{ \Gamma_{21} }{ \Gamma_{12} + \Gamma_{21} } L_{22} .
\end{align}
Note that $r_1$ represents the thermal equilibrium density matrix with $\braket{l_1,r_1} = \mathrm{tr}(r_1) = 1$ and $\mathcal{L}r_1 = 0$, as already discussed above.
It can be written in terms of the system Hamiltonian $H$ in its standard form as $r_1 = e^{-\beta H} /\mathrm{tr}(e^{-\beta H})$ by using the detailed balance condition $\Gamma_{nm}/\Gamma_{mn} = e^{-\beta(E_n - E_m)}$ that was already introduced in Appendix \ref{Appendix:System-Bath}.

For the eigenvalue $\lambda_2 = - \Gamma_{12} - \Gamma_{21}$, the left and right eigenvectors are
\begin{align}
    l_2 \propto \begin{pmatrix}
        \Gamma_{21} \\ -\Gamma_{12}
    \end{pmatrix} , 
    \qquad 
    r_2 \propto \begin{pmatrix}
        1 \\ -1
    \end{pmatrix} ,
\end{align}
which correspond to the normalized eigenoperators
\begin{align}
    l_2 = \frac{ \Gamma_{21} }{ \Gamma_{12} + \Gamma_{21} } L_{11} - \frac{ \Gamma_{12} }{\Gamma_{12} + \Gamma_{21} } L_{22} ,
    \qquad 
    r_2 = L_{11} - L_{22} .
    \label{eq:eigenoperatorsTLSlambda2}
\end{align}
The eigenoperators are orthonormal, $\braket{l_m,r_n} = \delta_{mn}$, and $\mathcal{L}r_m = \lambda_m r_m$ and $\mathcal{L}^\dag l_m = \lambda_m l_m$ holds by construction.

\twocolumngrid

	\onecolumngrid

\section{Frequency-dependent susceptibility}
\label{AppendixB}
The general susceptibility $\chi(t)$ in Eq.~\eqref{eq:total_susceptibility} of the main text can be split into four contributions:
A static contribution $\chi_\mathrm{st}(t)$
and the three contributions $\chi_{\alpha}(t)$, namely, Hamiltonian ($\alpha = H$), Sisyphus ($\alpha = \Gamma$), and Hermes ($\alpha = L$).
In the following, we will derive these contributions for a prepared initial system state with a diagonal density matrix $\rho = \sum_{m=1}^N p_m \ket{\phi_m}\bra{\phi_m}$.
In order to make use of the diagonalization of the unperturbed Lindblad superoperator discussed in Appendix \ref{section:diagonalizationSuperoperatorAppendix}, we decompose the three contributions in Eq.~\eqref{eq:SusceptContributionAppendix} into a linear combination of the basis operators $L_{mn} = \ket{\phi_m}\bra{\phi_n}$ such that
\begin{align}
    \mathcal{L}_\alpha'\rho = 
    \sum_{m,n=1}^N 
    \braket{\phi_m|\mathcal{L}_\alpha'\rho|\phi_n} L_{mn} .
\end{align}

\subsection{Hamiltonian contribution $\chi_{H}(t)$}
\label{Appendix:derivation_Hamiltonian}
In this case, 
Eq.~\eqref{eq:SusceptContributionAppendixHamiltonian} yields the matrix elements
\begin{align}
    \braket{\phi_m|\mathcal{L}_H'\rho|\phi_n}  
    &= 
    - i \braket{\phi_m|[ H' , \rho ]|\phi_n}
    =
    i (p_m-p_n) \braket{\phi_m| H'|\phi_n} .
\end{align}
Note that $\braket{\phi_m|\mathcal{L}_H'\rho|\phi_n} = 0$ for $m = n$, meaning that the operator $\mathcal{L}_H'\rho$ does not contain contributions from $L_{mm}$.
We can therefore make direct use of the eigenvalues $\lambda_{mn} = - \Gamma_{T_2}^{mn} - i (E_m-E_n)$ with the decoherence rates $\Gamma_{T_2}^{mn}$ in Eq.~\eqref{eq:general_Gamma_T_2}.
Then, the Hamiltonian susceptibility becomes
\begin{align}
    \chi_H(\omega) 
    &= \int_{-\infty}^\infty \chi_H(t) e^{i\omega t} dt 
    =  \int_{0}^\infty 
    \mathrm{tr}\{ A \, e^{t(i\omega + \mathcal{L})}  \mathcal{L}_H' \rho    \}
    dt
\nonumber \\
    &= 
    \sum_{\substack{m,n=1\\(m\neq n)}}^N
    \braket{\phi_m|\mathcal{L}_H'\rho|\phi_n} 
    \mathrm{tr}\{ A  L_{mn}   \}
    \int_{0}^\infty \, e^{-(\Gamma_{T_2}^{mn} + i (E_m-E_n-\omega))t} 
    dt 
\nonumber \\
    \label{eq:HamiltonianFinalAppendix}
    &= 
    \sum_{\substack{m,n=1\\(m\neq n)}}^N 
    (p_m-p_n) 
    \frac{ \braket{\phi_m|H'|\phi_n}
    \braket{\phi_n|A|\phi_m} }{E_m-E_n-\omega -i\Gamma_{T_2}^{mn}} ,
\end{align}
which is presented in Eq.~\eqref{eq:ULR_susceptibility-H} in the main text.
The Hamiltonian susceptibility is resonant at transition energies $\omega = E_m-E_n$ with a Lorentzian broadening due to the decoherence rates $\Gamma_{T_2}^{mn}$.

\subsection{Hermes contribution $\chi_{L}(t)$}
\label{Appendix:derivation_Hermes}
With the matrix elements of the derivative of the jump operators, 
\begin{align}
    \braket{\phi_m|L'_{ab}|\phi_n}
    = 
    \delta_{nb} \braket{\phi_m|\phi_a'} 
    -
    \delta_{ma} \braket{\phi_b|\phi_n'}  ,
\end{align}
a lengthy but straightforward simplification of Eq.~\eqref{eq:SusceptContributionAppendixHermes} yields the matrix elements
\begin{align}
    \braket{\phi_m|\mathcal{L}_L'\rho|\phi_n}  
    &= 
    \sum_{a,b = 1}^N \Gamma_{ab} \bra{\phi_m}\left( 
    L_{ab}' \rho L_{ab}^\dag 
+
L_{ab}\rho (L_{ab}^\dag)'
-
\frac{1}{2}
\left\{
(L_{ab}^\dag)' L_{ab}
+
L_{ab}^\dag
L_{ab}'
,\rho 
\right\} 
\right) \ket{\phi_n}
\nonumber \\
&= 
\sum_{a=1}^N \left( \Gamma_{na} p_a - \frac{1}{2} (p_m + p_n) \Gamma_{an} \right) \braket{\phi_m|L_{na}'|\phi_a}
+
\sum_{a=1}^N \left( \Gamma_{ma} p_a - \frac{1}{2} (p_m + p_n) \Gamma_{am} \right) \braket{\phi_a|L_{am}'|\phi_n}
\nonumber \\
&=
\braket{\phi_m|\phi_n'}
\Lambda_{mn},
\end{align}
with 
\begin{align}
\label{eq:LambdaCoefficientsHermes}
\Lambda_{mn} 
&= 
\sum_{a = 1}^N  
 \left( 
\frac{1}{2} (p_m + p_n)
(\Gamma_{am} - \Gamma_{an} )
- p_a ( \Gamma_{ma}  - \Gamma_{na} ) 
\right) 
= - (p_m - p_n ) \Gamma_{T_2}^{mn} + \mathcal{G}_m - \mathcal{G}_n ,
\end{align}
in terms of the decoherence rate $\Gamma_{T_2}^{mn}$ defined in Eq.~\eqref{eq:general_Gamma_T_2} and
\begin{align}
    \label{eq:thermalizationMeasureAppendix}
    \mathcal{G}_m = \sum_{a=1}^N ( \Gamma_{am} p_m - \Gamma_{ma} p_a ).
\end{align}
In general, $\mathcal{G}_m$ is a measure of the thermalization of the system.
For a completely thermalized state for which the populations $p_m$ are fully determined by the transition rates, the detailed balance condition $\Gamma_{mn} p_n = \Gamma_{nm} p_m$ results in $\mathcal{G}_m = 0$ and $\Lambda_{mn} = - (p_m - p_n ) \Gamma_{T_2}^{mn}$.

Since $\braket{\phi_m|\mathcal{L}_L'\rho|\phi_n} \propto \Lambda_{mn} = 0$ for $m = n$, the operator $\mathcal{L}_\alpha'\rho$ does not contain contributions from $L_{mm}$.
In the same way as for the Hamiltonian contribution, we can therefore make direct use of the eigenvalues $\lambda_{mn} = - \Gamma_{T_2}^{mn} - i (E_m-E_n)$ with the decoherence rates in Eq.~\eqref{eq:general_Gamma_T_2}.
Using the Hellmann-Feynman theorem, 
\begin{align}
    \label{eq:Hellman-Feynman}
    (E_m - E_n) \braket{\phi_m|\phi_n'} = \delta_{mn} E_m' - \braket{\phi_m|H'|\phi_n} ,
\end{align}
the Hermes susceptibility becomes
\begin{align}
    \chi_L(\omega) 
    &= \int_{-\infty}^\infty \chi_L(t) e^{i\omega t} dt 
    =  \int_{0}^\infty 
    \mathrm{tr}\{ A \, e^{t(i\omega + \mathcal{L})}  \mathcal{L}_L' \rho    \}
    dt
\nonumber \\
&=  \sum_{\substack{m,n=1\\(m\neq n)}}^N  \braket{\phi_m|\mathcal{L}_L' \rho|\phi_n}  \int_{0}^\infty 
    \mathrm{tr}\{ A \, e^{t(i\omega + \mathcal{L})} L_{mn}   \}
    dt
\nonumber \\
&=  \sum_{\substack{m,n=1\\(m\neq n)}}^N  \braket{\phi_m|\phi_n'}
\Lambda_{mn}  \mathrm{tr}\{ A L_{mn}   \}  \int_{0}^\infty 
    e^{-( \Gamma_{T_2}^{mn} + i (E_m-E_n-\omega)) t} 
    dt
\nonumber \\
    \label{eq:HermesFinalAppendix}
    &= \sum_{\substack{m,n=1\\(m\neq n)}}^N  
 \frac{i \Lambda_{mn} }{E_m - E_n} \, 
 \frac{ \braket{\phi_m|H'|\phi_n} \braket{\phi_n|A|\phi_m}  }{E_m-E_n-\omega -i \Gamma_{T_2}^{mn} } ,
\end{align}
which is the result presented in Eq.~\eqref{eq:HermesContributionMainText} in the main text. This result implements the Hermes contribution for arbitrary populations and it is insightful to separate the terms which will be vanishing for thermalized populations. This separation  reads $\chi_L=\chi_{L,\mathrm{DB}}+\chi_{L,\mathrm{BT}}$ with 
\begin{subequations}\label{eq:chiLDB_BT}
    \begin{align}
        \chi_{L,\mathrm{DB}}(\omega) &= -i\sum_{\substack{m,n=1\\(m\neq n)}}^N  
 \Gamma^{mn}_{T_2} \frac{p_m-p_n}{E_m - E_n} \, 
 \frac{ \braket{\phi_m|H'|\phi_n} \braket{\phi_n|A|\phi_m}  }{E_m-E_n-\omega -i \Gamma_{T_2}^{mn} }, 
 \\
 \chi_{L,\mathrm{BT}}(\omega) &=i\sum_{\substack{m,n=1\\(m\neq n)}}^N  
 \frac{\mathcal{G}_m-\mathcal{G}_n}{E_m - E_n} \, 
 \frac{ \braket{\phi_m|H'|\phi_n} \braket{\phi_n|A|\phi_m}  }{E_m-E_n-\omega -i \Gamma_{T_2}^{mn} }, 
    \end{align}
\end{subequations}
from which we identify the contribution present for detailed balance (DB) and its corrections beyond thermalization (BT). Notably, $\chi_{L,\mathrm{DB}}$ is identical to the full Hermes contribution derived in Ref.~\cite{peri_unified_2024}.

\subsection{Sisyphus contribution $\chi_{\Gamma}(t)$}
\label{Appendix:derivation_Sisyphus_corrected}
In this case, Eq.~\eqref{eq:SusceptContributionAppendixSisyphus} yields the matrix elements
\begin{align}
    \braket{\phi_m|\mathcal{L}_\Gamma'\rho|\phi_n}  
    &= 
    \sum_{a,b = 1}^N 
    \Gamma_{ab}'
    \bra{\phi_m}\left( 
    L_{ab} \rho L_{ab}^\dag  
    -\frac{1}{2}
    \left\{L^\dagger_{ab}L_{ab},\rho \right\} 
    \right)\ket{\phi_n}
= 
    \delta_{mn} 
    \sum_{\substack{a,b=1\\(a\neq b)}}^N 
    \Gamma'_{ab} p_b
    ( \delta_{am} - \delta_{bm} ) ,
\end{align}
such that
\begin{align}
    \mathcal{L}_\Gamma'\rho 
    = 
    \sum_{\substack{m,n=1\\(m\neq n)}}^N 
    \Gamma'_{mn} p_n 
    (L_{mm} - L_{nn}).
\end{align}
Note that dephasing rates $\Gamma_{mm}$ do not enter the Sisyphus contribution.
Since $\mathcal{L}_\Gamma'\rho$ is a superposition of all $L_{mm}$, we need to decompose it into a linear combination of right eigenoperators $r_k$ of the unperturbed Lindblad superoperator $\mathcal{L}$, as discussed in Appendix \ref{section:diagonalizationSuperoperatorAppendixMequalN}, 
with $\mathcal{L} r_k = \lambda_k r_k$ and real eigenvalues $\lambda_k \leq 0$.
This decomposition reads
\begin{align}
    \mathcal{L}_\Gamma'\rho 
&= 
    \sum_{k=1}^N \braket{l_k,\mathcal{L}_\Gamma'\rho } r_k
    =
    \sum_{\substack{k,m,n=1\\(m\neq n)}}^N 
    \Gamma_{mn}' p_n 
    \left( 
    \mathrm{tr}\{l_k^\dag L_{mm}\} 
    -
    \mathrm{tr}\{l_k^\dag L_{nn}\} 
    \right) 
    r_k
    \nonumber \\
&=
\sum_{\substack{k,m,n=1\\(m\neq n)}}^N
    \Gamma_{mn}' p_n 
    \left( 
    \braket{\phi_m|l_k^\dag|\phi_m} 
    -
    \braket{\phi_n|l_k^\dag|\phi_n} 
    \right) 
    r_k
    ,
\end{align}
with left eigenoperators $l_k$ that satisfy $\mathcal{L}^\dag l_k = \lambda_k l_k$ and the Hilbert-Schmidt inner product $\braket{A,B} = \mathrm{tr}(A^\dag B)$.
Then, the Sisyphus susceptibility becomes
\begin{align}
    \chi_\Gamma(\omega) 
    &= \int_{-\infty}^\infty \chi_\Gamma(t) e^{i\omega t} dt 
    =  
    \int_{0}^\infty 
    \mathrm{tr}\{ A \, e^{t(i \omega + \mathcal{L})}  \mathcal{L}_\Gamma' \rho \}
    dt
\nonumber \\
    &=  
    \sum_{\substack{k,m,n=1\\(m\neq n)}}^N
    \Gamma_{mn}' p_n 
    \left( 
    \braket{\phi_m|l_k^\dag|\phi_m} 
    -
    \braket{\phi_n|l_k^\dag|\phi_n} 
    \right) \mathrm{tr}\{ Ar_k \} 
    \int_{0}^\infty 
    e^{(\lambda_k + i\omega )t} 
    dt. 
\end{align}
The unique eigenvalue $\lambda_k = 0$ does not contribute to the summation as it comes with a left eigenoperator $l_k = \mathbbm{1}$, turning the term in brackets equal to zero.
All other eigenvalues are negative and nonzero. 
Furthermore, all left and right eigenoperators $l_k$ and $r_k$, respectively, are linear combinations of $L_{mm}$ with real coefficients, rendering them hermitian operators, $l_k^\dag = l_k$ and $r_k ^\dag = r_k$. 
This results in diagonal matrices with matrix elements, e.g., $\braket{\phi_m|r_k|\phi_n} = \delta_{mn} \braket{\phi_m|r_k|\phi_m}$, with which we finally get the Sisyphus susceptibility
\begin{align}
    \label{eq:SisyphusFinalAppendix}
    \chi_\Gamma(\omega) 
    &= 
    \sum_{\substack{k,l,m,n=1\\(m\neq n, \lambda_k < 0)}}^N  
    c_{k,lmn} \, \Gamma_{mn}' \, p_n \, 
    \frac{ i  \braket{\phi_l|A|\phi_l}  }{\omega-i\lambda_k} 
    ,
\end{align}
where we defined the expansion coefficients
\begin{align}
    \label{eq:sisyphus-coefficient}
    c_{k,lmn}
    =
    \braket{\phi_l|r_k|\phi_l}
    \Bigl( 
    \braket{\phi_m|l_k|\phi_m} 
    -
    \braket{\phi_n|l_k|\phi_n} 
    \Bigr)   .
\end{align}
This is the result presented in Eq.~\eqref{eq:Sisyphus solution} in the main text. To compare to previous approaches which only considered thermalized populations \cite{trif_dynamic_2018,peri_unified_2024}, a separation $\chi_\Gamma=\chi_{\Gamma,\mathrm{DB}}+\chi_{\Gamma,\mathrm{BT}}$ into detailed balance and beyond thermalization terms reads
\begin{subequations}
\label{eq:Sisyphus_DBBT}
    \begin{align}
        \chi_{\Gamma,\mathrm{DB}}(\omega) 
    &= -
    \sum_{\substack{k,l,m,n=1\\(m\neq n,\lambda_k < 0)}}^N
    c_{k,lmn} \, \Gamma_{mn} \, p_n' \, 
    \frac{ i  \braket{\phi_l|A|\phi_l}}{\omega-i\lambda_k},\\
   \chi_{\Gamma,\mathrm{BT}}(\omega) 
    &= -\sum_{\substack{k,l,m=1\\(\lambda_k < 0)}}^N
\braket{\phi_l|r_k|\phi_l}\braket{\phi_m|l_k|\phi_m} \, \mathcal{G}_m' \, 
    \frac{ i  \braket{\phi_l|A|\phi_l}  }{\omega-i\lambda_k}.
    \end{align}
\end{subequations}
Interestingly, in $\chi_{\Gamma,\mathrm{DB}}$ the derivative has moved from the rates to the populations which connects to the diagonal susceptibility of Ref.~\cite{trif_dynamic_2018}. Also, in $\chi_{\Gamma,\mathrm{BT}}$ the thermalization measure $\mathcal{G}_m$ defined in Eq.~\eqref{eq:thermalizationMeasureAppendix} emerges again. For our examples of either thermalized populations ($\mathcal{G}_m=0$) or constant populations ($p_n'=0$), one of the two contributions in Eq.~\eqref{eq:Sisyphus_DBBT} is always zero while the other produces the full Sisyphus susceptibility.

As an example, we give the explicit formula of the expansion coefficients for a two-level system $N = 2$. As discussed in Appendix \ref{sec:appendixExampleTLS}, the eigenvalues of the Lindblad dissipation matrix are $\lambda_1 = 0$ and $\lambda_2 = - \Gamma_{12} - \Gamma_{21}$.
Since $c_{1,lmn} = 0$, we only need the left and right eigenoperators for $\lambda_2$, which are defined in Eq.~\eqref{eq:eigenoperatorsTLSlambda2}.
This results in the coefficient
\begin{align}
    \label{eq:sisyphus_coefficient_TLS}
    c_{k,lmn}
    =
    \frac{  
    \Gamma_{21} 
    (\delta_{m1} - \delta_{n1})
    -
    \Gamma_{12}  (\delta_{m2} - \delta_{n2})
     }{ \Gamma_{12} + \Gamma_{21} } ( \delta_{l1} - \delta_{l2}) \delta_{k2}
    =
    (-1)^{l+m+2n} (1-\delta_{mn}) \delta_{k2} .
\end{align}
For general systems with $N>2$ energy levels, the precise form of the coefficient $c_{k,lmn}$ will be quite cumbersome.

\twocolumngrid
    \onecolumngrid
\section{Relation between susceptibility and band curvature}
\label{Appendix:Connection to curvature}
There are several limits in which the Lindblad-based linear response theory can recover the curvature of the energy bands.
Even if the environment is influencing system response dynamics, Hermes contributions compensate decoherence effects on Hamiltonian contributions, as we will discuss in the following. 

\subsection{Band curvature from low-frequency spectroscopy of isolated quantum systems}
\label{sec:lowFreqSpecIsoSysAppendix}
For isolated systems, $\Gamma_{mn} = 0$, both Sisyphus and Hermes contributions are zero, and the total susceptibility reduces to $\chi(\omega) = \chi_\mathrm{st} + \chi_H(\omega)$.
In the dispersive frequency regime, $\omega \ll |E_m-E_n|$, we get
\begin{align}
    \label{eq:ChiIsolatedLowFreqExpansionAppendix}
    \chi_H(\omega) = 
    \sum_{\substack{m,n=1\\(m\neq n)}}^N (p_m-p_n) 
    \frac{ \braket{\phi_m|H'|\phi_n}
    \braket{\phi_n|A|\phi_m} }{E_m-E_n} 
    + \mathcal{O}(\omega).
\end{align}
In cases where the observable $A = H'$ is the change in the Hamiltonian, and by applying the derivative $\partial_f$ again to the Hellman-Feynman theorem in Eq.~\eqref{eq:Hellman-Feynman} resulting in 
\begin{align}
    \label{eq:Hellman-Feynman-2nd}
     E_m'' 
     = 
     \braket{\phi_m|H''|\phi_m} 
     + 
     2 
     \sum_{\substack{n=1\\(n\neq m)}}^N 
     \frac{|\braket{\phi_m|H'|\phi_n}|^2}{E_m-E_n} ,
\end{align}
we obtain
\begin{align}
    \chi(\omega) 
    &= \sum_{m=1}^N p_m \braket{\phi_m|H''|\phi_m} 
    +
    \sum_{\substack{m,n=1\\(m\neq n)}}^N (p_m-p_n) 
    \frac{ |\braket{\phi_m|H'|\phi_n}|^2 }{E_m-E_n} 
    + \mathcal{O}(\omega)
    \nonumber \\
    &=
    \sum_{m=1}^N p_m E_m''
    -  
    \underbrace{ \sum_{\substack{m,n=1\\(m\neq n)}}^N (p_m + p_n) \frac{|\braket{\phi_m|H'|\phi_n}|^2}{E_m-E_n} }_{= \, 0}
    + \mathcal{O}(\omega)
    =
    \sum_{m=1}^N p_m E_m'' + \mathcal{O}(\omega) 
    \label{eq:TotalSuscpetIsolatedLowFreq}.
\end{align}
Since the second term vanishes under the summation (symmetric times antisymmetric in $m$ and $n$), the total zero-frequency susceptibility $\chi(0)$ reduces to a weighted sum of curvatures of each energy level, which is the result presented in Eq.~\eqref{eq:curvature}. 
This connection was inspired by Ref.~\cite{park_adiabatic_2020}. 
Therefore, the zero-frequency total susceptibility automatically reduces to this result whenever environmental interactions become negligible. 
In the case in which the modulated parameter $f = \mu$ is the chemical potential of the system, this curvature is called quantum capacitance $C_Q$ \cite{van_loo_single-shot_2026, liu_quantum_2026,zhang_gate_2025, kurilovich_microwave_2021, boutin_predictive_2025, peri_unified_2024,kitsenko_reflections_2026,persson_fast_2010,lambert_quantum_2016,malinowski_quantum_2022,jennings_probing_2025}.
In the case in which the modulated parameter $f = \phi$ is the superconducting phase difference or flux of the system, this curvature is called inverse Josephson induction $L^{-1}_J$ \cite{paila_current-phase_2009,baumgartner_josephson_2021,chiodi_probing_2011,trif_dynamic_2018}.


\subsection{Band curvature from low-frequency spectroscopy of open quantum systems}
\label{sec:dispersive_limit}
In more general setups in which interactions with the environment are not negligible, the two additional contributions to the total susceptibility, namely Sisyphus and Hermes contributions, can become relevant. 
In general, due to their similarities, the Hamiltonian and Hermes contributions in Eqs.~\eqref{eq:HamiltonianFinalAppendix} and \eqref{eq:HermesFinalAppendix} sum up to to the exact result
\begin{align}
    \chi_H(\omega)  + \chi_L(\omega)
    &=   
    \sum_{\substack{m,n=1\\(m\neq n)}}^N  
    \frac{p_m-p_n}{E_m - E_n} \
    \frac{ E_m - E_n - i \Gamma_{T_2}^{mn} }{ E_m - E_n - \omega -i \Gamma_{T_2}^{mn} } \braket{\phi_m|H'|\phi_n} \braket{\phi_n|A|\phi_m}+\chi_{L,\mathrm{BT}}(\omega)
    .\label{eq:Ramon_form}
\end{align}
While the first term has already been derived with a simplified approach for $A = H'$ and populations $p_m$ in thermal equilibrium (i.e., $\mathcal{G}_m = 0$) in the context of flux spectroscopy with $f = \phi$ \cite{trif_dynamic_2018}, the second term, $\chi_{L,\mathrm{BT}}$, represents corrections that appear when the populations $p_m$ of the initial state do not follow detailed balance.

In the dispersive frequency limit, $\omega \ll |E_m-E_n|$, the general expression reduces to 
\begin{align}
    \chi_H(\omega)  + \chi_L(\omega)
    &=   
    \sum_{\substack{m,n=1\\(m\neq n)}}^N  
    \frac{p_m-p_n}{E_m - E_n} \braket{\phi_m|H'|\phi_n} \braket{\phi_n|A|\phi_m}
    +
    \sum_{\substack{m,n=1\\(m\neq n)}}^N 
    \frac{i (\mathcal{G}_m - \mathcal{G}_n)}{E_m - E_n} 
    \
    \frac{ \braket{\phi_m|H'|\phi_n} \braket{\phi_n|A|\phi_m}}{ E_m - E_n -i \Gamma_{T_2}^{mn} } 
    +
    \mathcal{O}(\omega)
    .
\end{align}
Note that the first term is independent from the environmental decoherence rate $\Gamma_{T_2}^{mn}$ and represents the same result as for the Hamiltonian term in isolated systems, cf.~Eq.~\eqref{eq:ChiIsolatedLowFreqExpansionAppendix}, regardless of the interactions with the environment. The second term adds corrections beyond fully thermalized populations, as discussed before. Due to symmetry in indices, it can be shown to always be real valued.

Therefore, in cases where the observable $A = H'$ is the change in the Hamiltonian and by using Eq.~\eqref{eq:Hellman-Feynman-2nd}, we get
\begin{align}
    \chi_\mathrm{st} + \chi_H(\omega)  + \chi_L(\omega)
    &=   
    \sum_{m=1}^N p_m E_m''
    -
    \sum_{\substack{m,n=1\\(m\neq n)}}^N 
    \frac{ \mathcal{G}_m - \mathcal{G}_n }{E_m - E_n} 
    \
    \frac{  \Gamma_{T_2}^{mn} \ |\braket{\phi_m|H'|\phi_n}|^2 }{ (E_m - E_n)^2 + (\Gamma_{T_2}^{mn})^2 } 
    +
    \mathcal{O}(\omega)
    ,
    \label{eq:dispersiveLimitHamiltonianAndHermesAppendix}
\end{align}
where we have used that the imaginary part of the second term is zero due to the antisymmetry with respect to the indices $m$ and $n$. 
The first term recovers again the curvature result of the isolated system when combined with the static susceptibility, cf.~Eq.~\eqref{eq:TotalSuscpetIsolatedLowFreq}.
The completely thermalized case for $\mathcal{G}_m = 0$ is presented in Eq.~\eqref{eq:dispersiveLimitHamiltonianAndHermesMainText} in the main text. 

\subsection{Band curvature from spectroscopy of open quantum systems with strong dephasing}
\label{sec:dephasing_limit}
Hermes and Hamiltonian contributions are very similar when comparing the results in Eqs.~\eqref{eq:HamiltonianFinalAppendix} and \eqref{eq:HermesFinalAppendix}.
In the limit in which the environment only causes strong dephasing, $\Gamma_{mm} \gg \Gamma_{mn}$ $(m\neq n)$, the decoherence rate in Eq.~\eqref{eq:general_Gamma_T_2} becomes $\Gamma_{T_2}^{mn} \approx (\Gamma_{mm} +\Gamma_{nn}) / 2$, and, hence, $\Lambda_{mn}  \approx - (p_m - p_n ) \Gamma_{T_2}^{mn}$ in Eq.~\eqref{eq:LambdaCoefficientsHermes}.
While the Hamiltonian susceptibility becomes negligible, the Hermes contribution results in
\begin{align}
    \chi_L(\omega) 
    &=  \sum_{\substack{m,n=1\\(m\neq n)}}^N 
    (p_m - p_n ) 
    \frac{  \braket{\phi_m|H'|\phi_n} \braket{\phi_n|A|\phi_m} }{E_m - E_n} + \mathcal{O}\left(\frac{1}{\Gamma_{mm}}\right),
\end{align}
which is the same as the Hamiltonian susceptibility of the isolated system in the low-frequency regime, given in Eq.~\eqref{eq:ChiIsolatedLowFreqExpansionAppendix}. 
Hence, similar to Eq.~\eqref{eq:TotalSuscpetIsolatedLowFreq}, the total susceptibility becomes
\begin{align}
    \chi(\omega) = \sum_{m=1}^N p_m E_m'' + \chi_\Gamma(\omega) + \mathcal{O}\left(\frac{1}{\Gamma_{mm}}\right).
\end{align}
We see that the Hermes term compensates the decoherence effects on the Hamiltonian term in the strong dephasing limit, which results in the reappearance of the weighted sum of energy level curvatures. 
Therefore, one can also call it the semiclassical limit \cite{peri_unified_2024}. 
Up to the Sisyphus term, this is the result presented in Eq.~\eqref{eq:dephasing_limit} in the main text.

\twocolumngrid
	
\onecolumngrid
\section{Admittance and its relation to different parameter fluctuations}
\label{AppendixD}
\noindent To connect to the framework of circuit QED, it is necessary to express the linear response of the system in terms of the admittance $Y(\omega)$ of a circuit element.
In general, the admittance of such a circuit element is defined as the integral kernel of the current response $\delta I(t)$ through the element generated by a small applied voltage perturbation $\delta V(t)$, i.e., 
\begin{equation}
	\delta I(t)
    =
    \int_{-\infty}^t Y(t-\tau) \, \delta V(\tau)\, d\tau.
	\label{eq:ulr_admittance in time domain}
\end{equation}
In the frequency domain, the Fourier-transformed admittance is complex-valued and results in
\begin{equation}
	Y(\omega)=\frac{\delta I(\omega)}{\delta V(\omega)} ,
	\label{eq:ulr_admittance}
\end{equation}
where $\delta V(\omega)$ and $\delta I(\omega)$ are the Fourier transformed voltage and current through a measured circuit element.  
Eq.~\eqref{eq:ulr_admittance in time domain} allows us to relate the admittance to a more general susceptibility $\chi$ if the perturbed parameter $\delta f(t)$ can be related to a voltage fluctuation $\delta V(t)$. 
Comparing this with the general linear-response formula in Eq.~\eqref{eq:appendixLinearResponseAchiF}, the current $I$ is the observable $A$.
In hybrid systems there are two natural choices for the driving parameters: The gate voltage $f = \mu$, which directly biases the systems energy levels, as well as the flux $f = \phi$ controlling the superconducting phase difference between the superconducting leads.

\subsection{Admittance for gate perturbations}
\label{Appendix:Admittance for gate perturbations}
The simplest way to characterize quantum dot devices is through the gate current response. 
A small applied time-dependent gate voltage $\delta V(t)$ induces a time-dependent shift of the onsite energy, 
$\delta\mu(t) = \alpha e \, \delta V(t)$. 
The coupling strength is determined by the elementary charge $e$ and the so-called lever arm $\alpha = C_{G}/(C_{G}+C_{S})$, where $C_{G}$ denotes the gate capacitance and $C_{S}$ the capacitance of the quantum dot to the surrounding environment \cite{peri_unified_2024}.
Under the perturbation $\mu \to \mu + \delta\mu(t)$, the Hamiltonian becomes
$H \to H + \Pi_\mu \, \delta\mu(t)$, where 
\begin{equation}
	\Pi_{\mu}=\frac{\partial H}{\partial\mu}
   \label{eq:dipole-operator}
\end{equation}
represents the dipole operator which determines the induced polarization charge $\alpha e \braket{\Pi_\mu}$ at the gate.
Moreover, the temporal change of this polarization charge determines the gate current $\delta I_{G} = \alpha e\partial_t\braket{\Pi_\mu}$. 
In the frequency domain, where $\partial_t\to-i\omega$, the gate current becomes directly proportional to the linear-response operator. 
This allows us to express the gate admittance as
\begin{equation}
	Y(\omega)
    =
    \frac{\delta I_{G}(\omega)}{\delta V(\omega)}
    =
    -i (\alpha e)^2 \omega \, 
    \chi(\omega),
\end{equation}
where $\chi(\omega) = \delta\langle \Pi_\mu\rangle (\omega) / \delta\mu(\omega) $ is the susceptibility for the observable $A = \Pi_\mu$ in Eq.~\eqref{eq:appendixLinearResponseAchiF} under gate perturbations $\delta f(t) = \delta\mu(t)$. 
This formulation naturally connects to second-order perturbation theory, where the response is governed by the squared dipole transition matrix elements.

\subsection{Admittance for flux perturbations}
\label{Appendix:Admittance for flux perturbations}
In superconducting systems, the magnetic flux through a superconducting loop controls the superconducting phase difference across a Josephson junction. 
Under the perturbation $\phi \to \phi + \delta\phi(t)$, the Hamiltonian becomes
$H \to H + \Phi_0 \, I_s \, \delta\phi(t)$, where 
\begin{equation}
	I_{s} = \frac{1}{\Phi_0} \frac{\partial H}{\partial\phi} 
    \label{eq:lin_response_op_flux}
\end{equation}
is the Josephson current operator and $\Phi_0=\hbar/(2e)$ is the flux quantum.
Now, the Josephson relation $\partial_t\phi(t)= V(t) / \Phi_0$
allows us to relate the phase perturbation $\phi(t)=\phi_0+\delta\phi(t)$ to a voltage perturbation $\delta V(t)$.
In the frequency domain, where $\partial_t \to - i \omega$, this relation becomes 
\begin{equation}
	\delta V(\omega) = -i\omega\Phi_0 \, \delta\phi(\omega).
	\label{eq:ulr_Josephson relation in frequency domain}
\end{equation} 
The flux admittance follows as
\begin{equation}
		Y(\omega)
        =
        \frac{\delta \langle I_{s}\rangle(\omega)}{\delta V(\omega)}
        = 
        \frac{i}{\omega\Phi_0^2} \chi(\omega)
        ,
\end{equation}
where $\chi(\omega) =  \delta \langle \Phi_0 I_{s}\rangle(\omega) / \delta\phi(\omega)$
is the susceptibility for the observable $A = \Phi_0 I_s = \partial_\phi H$ in Eq.~\eqref{eq:appendixLinearResponseAchiF} under phase perturbations $\delta f(t) = \delta\phi(t)$.

\twocolumngrid
    
\onecolumngrid
\section{Kitaev Chain - many body Hamiltonian}
\label{AppendixE}
\subsection{Minimal chain in parity sectors}
\noindent The effective Hamiltonian for a minimal Kitaev chain is separated into global parity sectors
\begin{align}
	H=\begin{pmatrix}
		0	&	\Delta		& 0&0\\
		\Delta	&	-\mu_1-\mu_2& 0&0\\
		0&0&-\mu_1&-t\\
		0&0&-t&-\mu_2
	\end{pmatrix}
\end{align}
in the many-body basis $\{\ket{00},\ket{11},\ket{10},\ket{01}\}$, where $\Delta$ denotes CAR and $t$ denotes ECT.
In the even particle number sector, the energies scale with the average onsite energy $\mu_1+\mu_2$, while the odd sector scales with their difference $\mu_1-\mu_2$. In gate spectroscopy, as performed, for instance, in Ref.~\cite{van_loo_single-shot_2026} where the common superconducting lead is driven to shift the onsite energy on both sites equally, the odd parity sector will form linear bands while the even ones have curvature. As shown in Appendix~\ref{Appendix:Connection to curvature}, the quantum capacitance $C_Q$ is a measure of the curvature of the energy levels. Therefore, a measurable difference in quantum capacitance can detect the global parity of the quantum device.

\subsection{Kitaev-Josephson junction (KJJ)}
\label{Appendix:KJJ_matrix}
By coupling two minimal Kitaev chains with the Josephson coupling Hamiltonian introduced in Ref.~\cite{pino_minimal_2024},
\begin{align}
	H_J
    =
    -t_J \, e^{i\phi/2} \, d_{L,2}^\dagger d_{R,1} 
    -t_J \, e^{-i\phi/2} \, d_{R,1}^\dagger d_{L,2} 
\end{align}
with the tunnel coupling $t_J$, 
we can define the full KJJ Hamiltonian 
\begin{align}
	H_{KJJ}=H_{L}\otimes \mathbbm{1}_R+ \mathbbm{1}_L\otimes H_{R}+H_J\,,
\end{align}
where $H_{L,R}$ are the MKC Hamiltonians of Eq.~\eqref{eq:min_chain_Hamil} with additional idices for the left and right chain, respectively. Assuming a common onsite energy $\mu$ across all four quantum dots, the global even parity sector of the KJJ Hamiltonian in the eigenbasis of the uncoupled minimal Kitaev chains, 
$\{\ket{O_-,O_-},\ket{O_-,O_+},\ket{O_+,O_-},\ket{O_+,O_+},\ket{E_-,E_-},\ket{E_-,E_+},\ket{E_+,E_-},\ket{E_+,E_+}\}$,  reads
\begin{align}
    H_{KJJ}=\begin{pmatrix}
        \tilde{H}_\text{odd}&V^\dagger\\
        V&\tilde{H}_\text{even}
    \end{pmatrix}\,,\label{eq:HKJJ_matrix_form}
\end{align}
where the diagonal entries are the many-body MKC eigenenergies
\begin{align}
\tilde{H}_{\nu}=\left(
\begin{array}{cccc}
 2 \epsilon _-^{(\nu)} & 0 & 0 & 0 \\
 0 & \epsilon _-^{(\nu)}+\epsilon _+^{(\nu)} & 0 & 0 \\
 0 & 0 & \epsilon _-^{(\nu)}+\epsilon _+^{(\nu)} & 0 \\
 0 & 0 & 0 & 2 \epsilon _+^{(\nu)} \\
\end{array}
\right)\,,\label{eq:KJJ-local-parity-subblocks}
\end{align}
for local odd and even parity $\nu=\{\text{odd},\text{even}\}$. The Josephson Hamiltonian couples local parities with
\begin{align}
\label{eq:appendixE2_V}
   \resizebox{\textwidth}{!}{$ V=\left(
\begin{array}{cccc}
 \frac{\Delta  t_J \cos \left(\frac{\phi }{2}\right)}{\epsilon } & -\frac{i \Delta  t_J \sin \left(\frac{\phi }{2}\right)}{\epsilon } & -\frac{i \Delta 
   t_J \sin \left(\frac{\phi }{2}\right)}{\epsilon } & \frac{\Delta  t_J \cos \left(\frac{\phi }{2}\right)}{\epsilon } \\
 \frac{1}{2} t_J \left(-\frac{2 \mu  \cos \left(\frac{\phi }{2}\right)}{\epsilon }-i \sin \left(\frac{\phi }{2}\right)\right) & \frac{1}{2} t_J \left(\cos
   \left(\frac{\phi }{2}\right)+\frac{2 i \mu  \sin \left(\frac{\phi }{2}\right)}{\epsilon }\right) & \frac{1}{2} t_J \left(\cos \left(\frac{\phi
   }{2}\right)+\frac{2 i \mu  \sin \left(\frac{\phi }{2}\right)}{\epsilon }\right) & \frac{1}{2} t_J \left(-\frac{2 \mu  \cos \left(\frac{\phi
   }{2}\right)}{\epsilon }-i \sin \left(\frac{\phi }{2}\right)\right) \\
 \frac{1}{2} t_J \left(-\frac{2 \mu  \cos \left(\frac{\phi }{2}\right)}{\epsilon }+i \sin \left(\frac{\phi }{2}\right)\right) & \frac{1}{2} t_J \left(-\cos
   \left(\frac{\phi }{2}\right)+\frac{2 i \mu  \sin \left(\frac{\phi }{2}\right)}{\epsilon }\right) & \frac{1}{2} t_J \left(-\cos \left(\frac{\phi
   }{2}\right)+\frac{2 i \mu  \sin \left(\frac{\phi }{2}\right)}{\epsilon }\right) & \frac{1}{2} t_J \left(-\frac{2 \mu  \cos \left(\frac{\phi
   }{2}\right)}{\epsilon }+i \sin \left(\frac{\phi }{2}\right)\right) \\
 -\frac{\Delta  t_J \cos \left(\frac{\phi }{2}\right)}{\epsilon } & \frac{i \Delta  t_J \sin \left(\frac{\phi }{2}\right)}{\epsilon } & \frac{i \Delta 
   t_J \sin \left(\frac{\phi }{2}\right)}{\epsilon } & -\frac{\Delta  t_J \cos \left(\frac{\phi }{2}\right)}{\epsilon } \\
\end{array}
\right)$}.
\end{align}
Note that $\epsilon=2\sqrt{\mu^2+\Delta^2}$ is the MKC energy splitting. The Hamiltonian features three distinct energy regimes, i.e., low, central, and high. 
The low-energy subspace $\{\ket{E_-,E_-},\ket{O_-,O_-}\}$ forms the basis of a local parity qubit, as defined in Eq.~\eqref{eq:qubit-space} in the main text.
Interestingly, any qubit interaction is suppressed at the flux sweetspot $\phi=\pi$.
The high-energy regime is a symmetric copy of the low-energy regime. In the central energy regime, four energy levels emerge, each being a combination of one ground state and one excited state MKC. 
Odd and even local parities are coupled via the Josephson Hamiltonian. As long as only a common onsite energy is considered, two degenerate energy modes will form at energy $-2\mu$ . One of these modes is a dark state  $H_{KJJ}\ket{D}=-2\mu\ket{D}$ with $\ket{D}=(\ket{O_-,O_+}-\ket{O_+,O_-})/\sqrt{2}$ because it is also an eigenstate of the dipole operator Eq.~\eqref{eq:dipole-operator} and the current operator Eq.~\eqref{eq:lin_response_op_flux} so that $\Pi_\mu\ket{D}=-2\mu\ket{D}$ and $I_s\ket{D}=0$. Therefore, all response elements vanish $|\bra{\psi_i}\Pi_\mu\ket{D}|=0=|\bra{\psi_i}I_s\ket{D}|$ for all other Hamiltonian eigenstates $\ket{\psi_i}$. The dark state fully decouples from the system and does not contribute to any gate or flux response.
In fact, one can show that, at $\phi=\pi$, the global even KJJ can be solved analytically by the eigenenergies 
\begin{subequations}\label{eq:KJJ_even_analytical}
\begin{align}
    E_1(\phi=\pi)&=-2\mu-\sqrt{\left(\mu+\frac{t_J}{2}\right)^2+\Delta^2}-\sqrt{\left(\mu-\frac{t_J}{2}\right)^2+\Delta^2},\\
    E_2(\phi=\pi)&=-2\mu-\sqrt{t_J^2+4t^2},\\
    E_3(\phi=\pi)&=-2\mu-\sqrt{\left(\mu+\frac{t_J}{2}\right)^2+\Delta^2}+\sqrt{\left(\mu-\frac{t_J}{2}\right)^2+\Delta^2},\\
    E_4(\phi=\pi)&=-2\mu,\\
    E_5(\phi=\pi)&=-2\mu,\\
    E_6(\phi=\pi)&=-2\mu+\sqrt{\left(\mu+\frac{t_J}{2}\right)^2+\Delta^2}-\sqrt{\left(\mu-\frac{t_J}{2}\right)^2+\Delta^2},\\
    E_7(\phi=\pi)&=-2\mu+\sqrt{t_J^2+4t^2},\\
    E_8(\phi=\pi)&=-2\mu+\sqrt{\left(\mu+\frac{t_J}{2}\right)^2+\Delta^2}+\sqrt{\left(\mu-\frac{t_J}{2}\right)^2+\Delta^2},
\end{align}
\end{subequations}
which are sums of MKCs shifted to $\pm t_J/2$. With this, the quantum capacitance of the lowest energy state reads exactly
\begin{align}
    C_{Q,1}(\phi=\pi)=-\frac{\Delta ^2 }{\left(\Delta ^2+\left(\mu +\frac{t_J}{2}\right)^2\right)^{3/2}}-\frac{\Delta ^2 }{\left(\Delta ^2+\left(\mu -\frac{t_J}{2}\right)^2\right)^{3/2}}.\label{eq:CQ_flux_sweetspot}
\end{align}
These are depicted in Fig.~\ref{fig:KJJ_curvature}a.

\subsection{Linear response operators}
To compute the linear response, we need the matrix forms of the Hamiltonian derivatives, namely $\partial_f H_{KJJ}$ for the gate and flux perturbations $f = \mu$ and $f = \phi$, respectively. For common gate spectroscopy, the dipole operator is first calculated in second quantization as 
\begin{equation}
    \partial_\mu H_{KJJ}=-\sum_{\substack{i=L,R\\j=1,2}}d^\dagger_{i,j}d_{i,j}\,.
\end{equation}
In the basis of decoupled MKCs, it reads 
\begin{align}
    \partial_\mu H_{KJJ}=\left(
\begin{array}{cccccccc}
 1 & 0 & 0 & 0 & 0 & 0 & 0 & 0 \\
 0 & 1 & 0 & 0 & 0 & 0 & 0 & 0 \\
 0 & 0 & 1 & 0 & 0 & 0 & 0 & 0 \\
 0 & 0 & 0 & 1 & 0 & 0 & 0 & 0 \\
 0 & 0 & 0 & 0 & \left(\frac{\mu }{\sqrt{\Delta ^2+\mu ^2}}+1\right)^2 & \frac{\Delta  \left(\sqrt{\Delta ^2+\mu ^2}+\mu
   \right)}{\Delta ^2+\mu ^2} & \frac{\Delta  \left(\sqrt{\Delta ^2+\mu ^2}+\mu \right)}{\Delta ^2+\mu ^2} & \frac{\Delta
   ^2}{\Delta ^2+\mu ^2} \\
 0 & 0 & 0 & 0 & \frac{\Delta  \left(\sqrt{\Delta ^2+\mu ^2}+\mu \right)}{\Delta ^2+\mu ^2} & \frac{\Delta ^2}{\Delta ^2+\mu
   ^2} & \frac{\Delta ^2}{\Delta ^2+\mu ^2} & \frac{\Delta  \left(\sqrt{\Delta ^2+\mu ^2}-\mu \right)}{\Delta ^2+\mu ^2} \\
 0 & 0 & 0 & 0 & \frac{\Delta  \left(\sqrt{\Delta ^2+\mu ^2}+\mu \right)}{\Delta ^2+\mu ^2} & \frac{\Delta ^2}{\Delta ^2+\mu
   ^2} & \frac{\Delta ^2}{\Delta ^2+\mu ^2} & \frac{\Delta  \left(\sqrt{\Delta ^2+\mu ^2}-\mu \right)}{\Delta ^2+\mu ^2} \\
 0 & 0 & 0 & 0 & \frac{\Delta ^2}{\Delta ^2+\mu ^2} & \frac{\Delta  \left(\sqrt{\Delta ^2+\mu ^2}-\mu \right)}{\Delta ^2+\mu
   ^2} & \frac{\Delta  \left(\sqrt{\Delta ^2+\mu ^2}-\mu \right)}{\Delta ^2+\mu ^2} & \left(\frac{\mu }{\sqrt{\Delta ^2+\mu
   ^2}}-1\right)^2 \\
\end{array}
\right)\,.
\end{align}
Notably, the local odd sector is an identity while the local even sector has a nontrivial structure. 
For flux spectroscopy, the Josephson current operator through the junction is determined in second quantization as 
\begin{equation}
    \partial_\phi H_{KJJ} = 
    - \frac{i t_J}{2} \, e^{i\phi/2} \, d_{L,2}^\dagger d_{R,1} 
    + \frac{i t_J}{2} \, e^{-i\phi/2} \, d_{R,1}^\dagger d_{L,2} .
\end{equation}
In the basis of decoupled MKCs, it reads
\begin{align}
    \partial_\phi H_{KJJ}=\begin{pmatrix}
        0&(\partial_\phi V)^\dagger\\
        \partial_\phi V&0
    \end{pmatrix},
\end{align}
with $V$ defined in Eq.~\eqref{eq:appendixE2_V}.
Because the flux couples the local odd and even parities, the block-diagonal is zero. The transformation into the decoupled MKC eigenspace does not depend on $\phi$. Therefore, the derivative can directly be applied on the matrix form of $H_{KJJ}$ in Eq.~\eqref{eq:HKJJ_matrix_form}.

\subsection{Low-energy regime }
\label{sec:perturbative_low_regime}
When considering the weak-coupling limit $t_J\ll|t+\Delta|/2$, all three energy regimes can be separated from each other. A projective treatment for the even parity low energy subspace has already been introduced in Ref.~\cite{pino_minimal_2024}. It revealed mirror symmetric flux energy bands around the MKC energy $E=\epsilon_-^\text{even}+\epsilon_-^\text{odd}$. However, the full spectrum does not feature such symmetry, see Fig.~\ref{fig:KJJ_spectra}c. Perturbative corrections can be computed with a Schrieffer-Wolff transformation (SWT). The analytical calculations were performed using SymPT \cite{reascos_universal_2025,diotallevi_sympt_2024}.
The effective low-energy Hamiltonian in the basis $\{\tilde{\ket{E_-}},\tilde{\ket{O_-}}\}$, which designates the qubit space, yields
\begin{align}
    H_\text{low}=\begin{pmatrix}
        2\epsilon_-^\text{even}-
        \frac{t_J^2\Delta^2\cos^2\frac{\phi}{2}}{\epsilon^2(2t+\epsilon)}
        -\frac{2t_J^2\Delta^2\sin^2\frac{\phi}{2}}{\epsilon^3}
        &t_J\frac{\Delta}{\epsilon}\cos\frac{\phi}{2}\\
        t_J\frac{\Delta}{\epsilon}\cos\frac{\phi}{2}
        &2\epsilon_-^\text{odd}-
        t_J^2 \left(\frac{\Delta^2}{\epsilon^2(2t+\epsilon)}
        +\frac{\mu^2}{t\epsilon^2} \right)\cos^2\frac{\phi}{2}
        -\frac{t_J^2}{4t}\sin^2\frac{\phi}{2}
    \end{pmatrix},
    \label{eq:QubitSWT}
\end{align}
where 
\begin{subequations}\label{eq:perturbation_basis}
\begin{align}
    \label{eq:perturbation_basis_even}\tilde{\ket{E_-}}
    &=|E_-,E_-\rangle
    +\frac{it_J\Delta}{\epsilon^2}
    \sin\frac{\phi}{2}
    \Big(|O_-,O_+\rangle+|O_+,O_-\rangle\Big)
    +\frac{t_J\Delta}{\epsilon(2t+\epsilon)}
    \cos\frac{\phi}{2}\,\ket{O_+,O_+},\\
    \tilde{\ket{O_-}}
    &=|O_-,O_-\rangle
    -\frac{\mu t_J}{2t\epsilon}
    \cos\frac{\phi}{2}
    \Big(\ket{E_+,E_-}+\ket{{E_-,E_+}}\Big)
    +\frac{i t_J}{4t}
    \sin\frac{\phi}{2}
    \Big(\ket{E_+,E_-}-\ket{{E_-,E_+}}\Big)
    -\frac{\Delta t_J}{\epsilon(2t+\epsilon)}
    \cos\frac{\phi}{2}
    \ket{E_+,E_+},\label{eq:perturbation_basis_odd}
\end{align}
\end{subequations}
are first-order corrections to the unperturbed qubit basis.
We will denote the qubit eigenstates with $\ket{\nicefrac{\uparrow}{\downarrow}}$. Up to second order in $t_J/|t+\Delta|$ their eigenenergies read
\begin{align}
        E^\text{even}_{\nicefrac{\uparrow}{\downarrow}} &= \epsilon_-^\text{even}+\epsilon_-^\text{odd}
        -\left(\frac{\Delta^2}{\epsilon^2(2t+\epsilon)}+\frac{\mu^2}{2\epsilon^2t}\right)t_J^2\cos^2\frac{\phi}{2}
        -\left(\frac{\Delta^2}{\epsilon^3}+\frac{1}{8t}\right)t_J^2\sin^2\frac{\phi}{2}\notag\\
        &\quad \pm\sqrt{\left(t-\frac{\epsilon}{2}-\frac{\mu^2}{2\epsilon^2t}t_J^2\cos^2\frac{\phi}{2}+\left(\frac{\Delta^2}{\epsilon^3}-\frac{1}{8t}\right)t_J^2\sin^2\frac{\phi}{2}\right)^2+t_J^2\frac{\Delta^2}{\epsilon^2}\cos^2\frac{\phi}{2}}.
\end{align}    
Corrections of higher energy regimes, i.e. $\mathcal{O}(t_J^2)$, lift the mirror symmetry between the first and second band around $E=\epsilon_-^\text{even}+\epsilon_-^\text{odd}$. This approximates the exact energy bands in Fig.~\ref{fig:KJJ_spectra} well. The two eigenenergies cross at $\phi=\pi$ only when tuned to the coupling sweetspot $\Delta=t$ and gate sweetspot $\mu=0$ as the diagonals become proportional to identity.
If the KJJ is probed at the flux sweet spot $\phi=\pi$, the basis states $\{\tilde{\ket{E_-}},\tilde{\ket{O_-}}\}$ are KJJ eigenstates up to first order in $t_J$ as Eq.~\eqref{eq:QubitSWT} becomes diagonal. These are the low energy states depicted in Fig.~\ref{fig:KJJ_spectra}b. Their quantum capacitances or gate curvatures then read
\begin{subequations}\begin{align}
    C_{Q,\downarrow}&=-\frac{16\Delta^2}{\epsilon^3}+96\,t_J^2\Delta^2\left(\frac{\Delta^2-4\mu^2}{\epsilon^7}\right)+\mathcal{O}(t_J^3),\\
    C_{Q,\uparrow}&=0.
\end{align}\end{subequations}
These quantum capacitances match the exact calculations in Eq.~\eqref{eq:CQ_flux_sweetspot} up to second order. The latter are depicted in Fig.~\ref{fig:KJJ_curvature}a. Interestingly, this coincides with the curvature sum of two minimal chains that are shifted  $\mu\to\mu\pm t_J/2$.
The flux curvatures or inverse Josephson induction $L^{-1}_{\nicefrac{\uparrow}{\downarrow}}=\partial_\phi E_{\nicefrac{\uparrow}{\downarrow}}^\text{even}$ at $\mu=0$ read
\begin{align}
    L^{-1}_{\nicefrac{\uparrow}{\downarrow}} =& -\frac{t_J^2}{16}\frac{ t^2+t\Delta+\Delta^2}{\Delta  t (\Delta +t)}\cos (\phi )\notag\\
    &\pm
    \frac{t_J^2 \left(\cos (\phi ) \left(-32 \Delta  t^3+t^2 \left(48 \Delta ^2+t_J^2\right)-2 \Delta  t
   \left(16 \Delta ^2+t_J^2\right)+\Delta ^2 t_J^2\right)-t_J^2 \cos (2 \phi ) (t-\Delta
   )^2\right)}{32 \Delta  t \sqrt{16 \Delta ^2 t^2 t_J^2 \cos ^2\left(\frac{\phi }{2}\right)+(t-\Delta )^2
   \left(t_J^2 \sin ^2\left(\frac{\phi }{2}\right)-16 \Delta  t\right)^2}}\label{eq:global_even_flux_curvature}\\
   &\mp\frac{t_J^4 \sin ^2(\phi )
   \left(t_J^2 \sin ^2\left(\frac{\phi }{2}\right) (t-\Delta )^2-8 \Delta  t \left(2 \Delta ^2+2 t^2-3 \Delta 
   t\right)\right)^2}{32 \Delta  t \left(16\Delta^2t^2 t_J^2 \cos ^2\left(\frac{\phi }{2}\right)+(t-\Delta )^2 \left(t_J^2 \sin^2 \left(\frac{\phi}{2} \right)-16 \Delta 
   t\right)^2\right)^{3/2}}\,.\notag
\end{align}
Influence of higher energy states on the qubit space thus causes an emerging asymmetry of ground and excited state, as seen in Fig.~\ref{fig:KJJ_curvature}b. The second and third line simplify to $\pm\frac{t_J}{8}\left|\cos\frac{\phi}{2}\right|$ for $\Delta=t$ and therefore become degenerate with the global odd Josephson induction we will derive in Eq.~\eqref{eq:global_odd_flux_curvature} below.
The analysis in the high energy regime is analogue with the perturbative basis $\{\tilde{\ket{E_+}},\tilde{\ket{O_+}}\}$.

\subsection{Central-energy regime}
\label{sec:perturbative_central_regime}
To analyze the central states qualitatively, we project Eq.~\eqref{eq:HKJJ_matrix_form} into the central energy regime subspace $\{\ket{O_-,O_+},\ket{O_+,O_-},\ket{E_-,E_+},\ket{E_+,E_-}\} $. The Hamiltonian reads
\begin{align}
    H_c=\begin{pmatrix}
        -2\mu   &   0 & 
        \frac{t_J}{2\epsilon} \eta^\ast(\phi) 
   & -\frac{t_J}{2\epsilon} \eta(\phi)  \\
    0   &   -2\mu & 
           \frac{t_J}{2\epsilon} \eta^\ast(\phi)  & -\frac{t_J}{2\epsilon} \eta(\phi) \\
    \frac{t_J}{2\epsilon} \eta(\phi)  
   &\frac{t_J}{2\epsilon} \eta(\phi) 
   &-2\mu &0\\
    -\frac{t_J}{2\epsilon} \eta^\ast(\phi)  
   &-\frac{t_J}{2\epsilon} \eta^\ast(\phi) 
   &0&-2\mu\\
    \end{pmatrix}
\end{align}
with the short-hand notation $\eta(\phi)=\epsilon\cos\frac{\phi}{2}+2i\mu\sin{\frac{\phi}{2}}$. It features two degenerate eigenvalues $\lambda_{1,2}=-2\mu$ and around them two symmetrical eigenvalues $\lambda_{3,4}=-2\mu\pm t_J|\eta(\phi)|/{\epsilon}$. The eigenstates are 
\begin{subequations}
\begin{align}
    \ket{\lambda_1}&=\ket{O_-,O_+}-\ket{O_+,O_-}\,,\\
    \ket{\lambda_2}&=\eta(\phi)\ket{E_-,E_+}
    +\eta^\ast(\phi)\ket{E_-,E_+},\\
    \ket{\lambda_{3,4}}&=\pm|\eta(\phi)|\ket{O_-,O_+}\pm|\eta(\phi)|\ket{O_+,O_-}
    +\eta(\phi)\ket{E_-,E_+}
    -\eta^\ast(\phi)\ket{E_-,E_+}.
\end{align} 
\end{subequations}
Note, that $\ket{\lambda_1}=\sqrt{2}\ket{D}$ is a dark state because it is also an eigenstate of the dipole and current operator.
In Fig.~\ref{fig:KJJ_spectra}\change{a}, the KJJ common gate spectrum is depicted at the flux sweet spot $\phi=\pi$. Therefore, eigenenergies read $\lambda_{1,2}=-2\mu$ and $\lambda_{3,4}=-2\mu\pm 2t_J\mu/{\epsilon}$, while the eigenstates simplify to 
\begin{subequations}
\begin{align}
    \ket{\lambda_1}&=\ket{O_-,O_+}-\ket{O_+,O_-}\,,\\
    \ket{\lambda_2}&=\ket{E_-,E_+}-\ket{E_-,E_+}\,,\\
    \ket{\lambda_{3,4}}&=\pm\ket{O_-,O_+}\pm\ket{O_+,O_-}
    +i\ket{E_-,E_+}
    +i\ket{E_-,E_+}\,.
\end{align}   
\end{subequations}
In Fig.~\ref{fig:KJJ_spectra}\change{b}, the KJJ flux spectrum is depicted in the gate sweetspot $\mu=0$. Therefore, eigenenergies read $\lambda_{1,2}=-2\mu$ and $\lambda_{3,4}=\pm t_J\cos\frac{\phi}{2}$ while the eigenstates simplify to 
\begin{subequations}\begin{align}
    \ket{\lambda_1}&=\ket{O_-,O_+}-\ket{O_+,O_-}\,,\\
    \ket{\lambda_2}&=\ket{E_-,E_+}+\ket{E_-,E_+}\,,\\
    \ket{\lambda_{3,4}}&=\pm\ket{O_-,O_+}\pm\ket{O_+,O_-}
    +\ket{E_-,E_+}
    -\ket{E_-,E_+}\,.
\end{align}   
\end{subequations}

\subsection{Global odd KJJ}
\label{sec:global_odd_KJJ}
\begin{figure}
    \centering
    \includegraphics[width=0.5\linewidth]{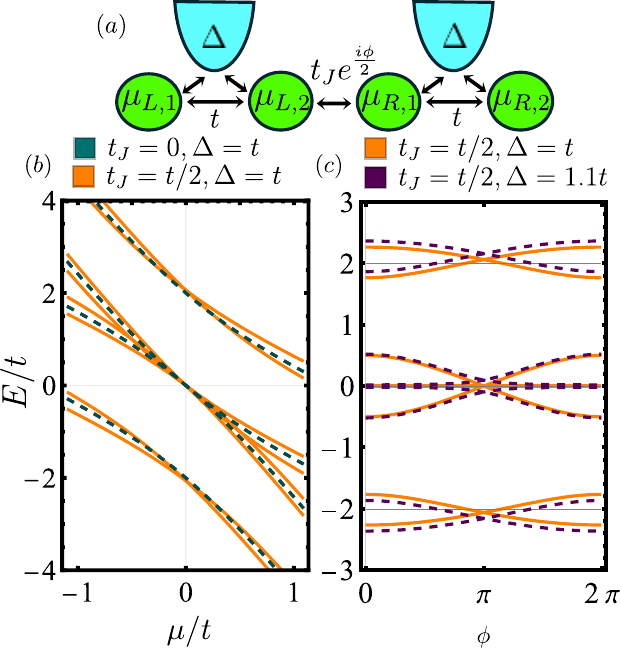}
    \caption{\textbf{(a)} Two minimal Kitaev chains tunnel coupled with $t_J$ form a Kitaev-Josephson Junction (KJJ). The superconducting leads have a phase difference of $\phi$. \textbf{(b)} Common gate spectrum in comparison for decoupled (green) and coupled (orange) KJJ in odd global parity at $\phi=\pi$ and $\Delta=t$. \textbf{(c)} Flux spectrum comparing sweetspot $\Delta=t$ (orange) with a detuned $\Delta= 1.1t$ (purple) system at $\mu=0$.}
    \label{fig:KJJ_ODD}
\end{figure}
In the global odd parity sector, the Hamiltonian in the basis $$\{\ket{O_-,E_-},\ket{O_-,E_+},\ket{O_+,E_-},\ket{O_+,E_+},\ket{E_-,O_-},\ket{E_+,O_-},\ket{E_-,O_+},\ket{E_+,O_+}\}$$ reads
\begin{align}
    H_{KJJ}^\text{odd}=\begin{pmatrix}
        \tilde{H}&U^\dagger\\
        U&\tilde{H}
    \end{pmatrix},
\end{align}
where
\begin{align}
\tilde{H}=\begin{pmatrix}
    \epsilon_-^\text{even}+\epsilon_-^\text{odd}&0&0&0\\
    0&\epsilon_+^\text{even}+\epsilon_-^\text{odd}&0&0\\
    0&0&\epsilon_-^\text{even}+\epsilon_+^\text{odd}&0\\
    0&0&0&\epsilon_+^\text{even}+\epsilon_+^\text{odd}
\end{pmatrix},
\end{align}
and
\begin{align}
U=\frac{t_J}{\epsilon}\left(
\begin{array}{cccc}
 -\frac{1}{2} \eta ^*(\phi ) & -i \Delta  \sin \left(\frac{\phi }{2}\right) & -\frac{1}{2} i \eta ^*(\phi +\pi ) & -\Delta  \cos \left(\frac{\phi }{2}\right) \\
 -i \Delta  \sin \left(\frac{\phi }{2}\right) & -\frac{1}{2}\eta (\phi ) & \Delta  \cos \left(\frac{\phi }{2}\right) & \frac{1}{2} i \eta (\phi +\pi ) \\
 -\frac{1}{2} i \eta ^*(\phi +\pi ) & -\Delta  \cos \left(\frac{\phi }{2}\right) & \frac{1}{2}\eta ^*(\phi ) & i \Delta  \sin \left(\frac{\phi }{2}\right) \\
 \Delta  \cos \left(\frac{\phi }{2}\right) & -\frac{1}{2} i \eta (\phi +\pi ) & -i \Delta  \sin \left(\frac{\phi }{2}\right) & \frac{1}{2} \eta (\phi )\\
\end{array}
\right).
\end{align}
The energy spectra for common gate and flux perturbations are depicted in Fig.~\ref{fig:KJJ_ODD}.
Using a SWT to separate high, low, and central energy blocks up to second order in $t_J$, we find the effective low energy Hamiltonian 
\begin{equation}
    H_\text{low}^\text{odd}
    =-\left(2\mu+t+\frac{\epsilon}{2}+
    \frac{t_J^2\Delta^2\cos^2{\frac{\phi}{2}}}{\epsilon^2(2t+\epsilon)}+
    \frac{t_J^2\Delta^2\sin^2{\frac{\phi}{2}}}{\epsilon^3}+
    \frac{t_J^2|\eta(\phi+\pi)|^2}{8t\epsilon^2}\right)\mathbbm{1}-
    \frac{t_J}{2\epsilon}\left(\Real(\eta(\phi))\sigma_x+\Imag(\eta(\phi))\sigma_y\right) ,
\end{equation}
where the Pauli matrices $\sigma_{x,y}$ and identity $\mathbbm{1}$ are defined in the basis $\{\ket{0_\text{odd}},\ket{1_\text{odd}}\}$ with
\begin{subequations}\begin{align}
    \ket{0_\text{odd}}&=\ket{O_-,E_-}
    -i\frac{t_J\Delta}{\epsilon^2}\sin{\frac{\phi}{2}}\ket{E_+,O_-}
    -i\frac{t_J\eta^\ast(\phi+\pi)}{4t\epsilon}\ket{E_-,O_+}
    +\frac{\Delta t_J\cos\frac{\phi}{2}}{\epsilon(2t+\epsilon)}\ket{E_+,O_+},\\
    \ket{1_\text{odd}}&=\ket{E_-,O_-}
    +i\frac{t_J\Delta}{\epsilon^2}\sin{\frac{\phi}{2}}\ket{O_-,E_+}
    +i\frac{t_J\eta(\phi+\pi)}{4t\epsilon}\ket{O_+,E_-}
    -\frac{\Delta t_J\cos\frac{\phi}{2}}{\epsilon(2t+\epsilon)}\ket{O_+,E_+}.
\end{align}\end{subequations}
Interestingly, $H_\text{low}^\text{odd}$ has no $\sigma_z$ contribution. Thus, the eigenenergies are
\begin{align}
    E_{\nicefrac{\uparrow}{\downarrow}}^\text{odd}=-\left(2\mu+t+\frac{\epsilon}{2}+
    \frac{t_J^2\Delta^2\cos^2{\frac{\phi}{2}}}{\epsilon^2(2t+\epsilon)}+
    \frac{t_J^2\Delta^2\sin^2{\frac{\phi}{2}}}{\epsilon^3}+
    \frac{t_J^2|\eta(\phi+\pi)|^2}{8t\epsilon^2}\right)\pm\frac{t_J}{2\epsilon}|\eta(\phi)|+\mathcal{O}(t_J^3).
    \label{eq:odd_low_regime_energies}
\end{align}
At the flux sweet spot $\phi=\pi$, the qubit eigenenergies can be solved exactly, $ E^\text{odd}_{\nicefrac{\uparrow}{\downarrow}}=-2\mu-t-\sqrt{\Delta^2+(\mu\pm t_J/2)^2}$, and they are equal to the even low-energy bands of an MKC with shifted $\mu \to \mu \pm t_J/2$ and reduced in total energy by $\mu+t$. The quantum capacitance then reads
\begin{align}
    C_{Q,{\nicefrac{\uparrow}{\downarrow}}}^\text{odd}=-\frac{\Delta^2}{\left(\Delta ^2+\left(\mu\pm \frac{t_J}{2}\right) ^2\right)^{3/2}}.
\end{align}
The exact curvatures are depicted in Fig.~\ref{fig:KJJ_curvature}a. Inverse Josephson induction $L_{\nicefrac{\uparrow}{\downarrow}}^{-1}=\partial_\phi^2 E_{\nicefrac{\uparrow}{\downarrow}}^\text{odd}$ at $\mu=0$ yields
\begin{align}
    L_{\nicefrac{\uparrow}{\downarrow}}^{-1}=
    -\frac{t_J^2}{16}\frac{ t^2+t\Delta+\Delta^2}{\Delta  t (\Delta +t)}\cos (\phi )\pm
    \frac{t_J}{8}\left|\cos{\frac{\phi}{2}}\right|\label{eq:global_odd_flux_curvature}.
\end{align}
The first term of Eq.~\eqref{eq:global_odd_flux_curvature} is identical to the first term of the global even parity inverse Josephson induction Eq.~\eqref{eq:global_even_flux_curvature}. However, the second term is independent of couplings $t$ or $\Delta$ and equal to global even parity only when $\Delta=t$. That is, where global parity eigenenergies are degenerate.

\twocolumngrid
    
\onecolumngrid
\section{Additional susceptibility measurements}
\label{AppendixF}

\subsection{KJJ contributons in common gate susceptibility}
\label{Appendix:Susceptibility_contributions_KJJ_gate}
\begin{figure}
    \centering
    \includegraphics[width=\linewidth]{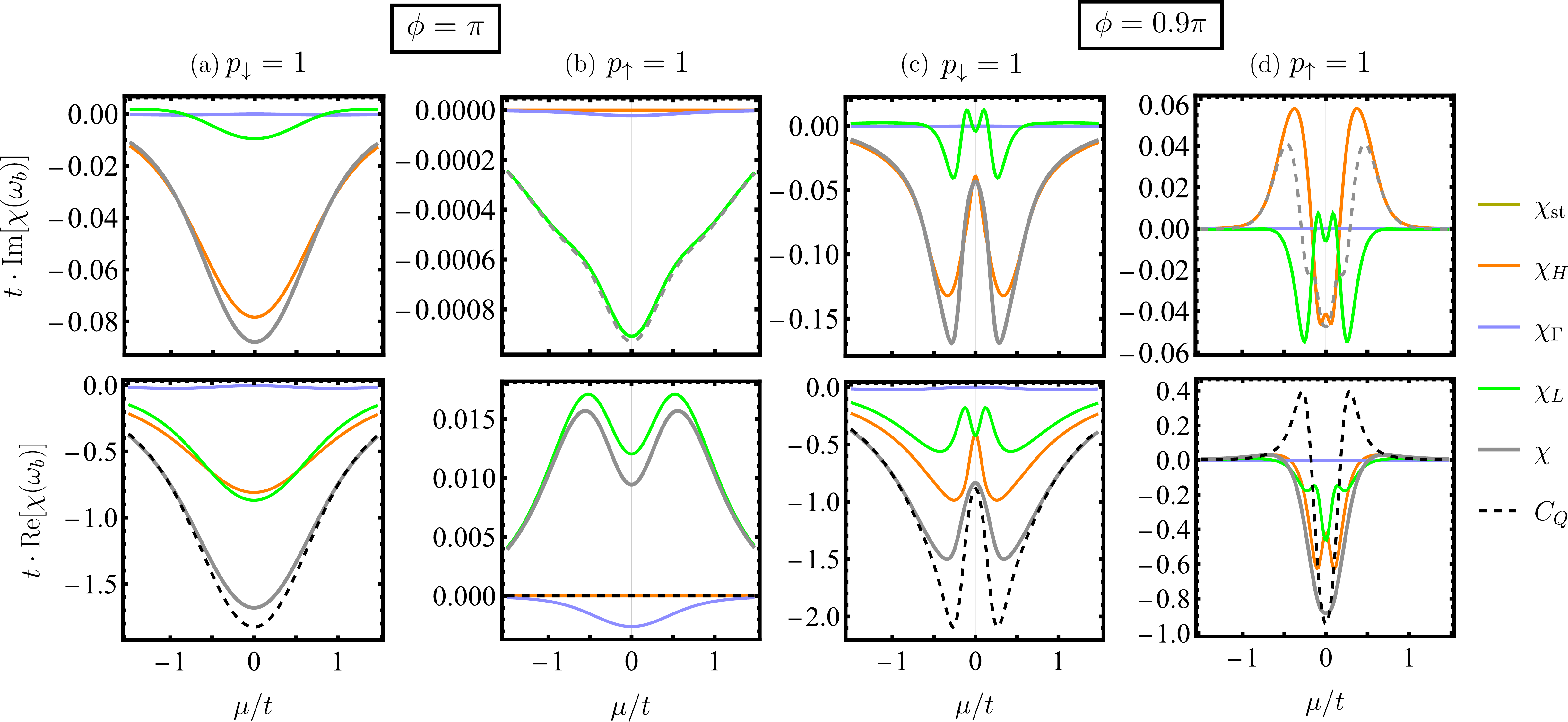}
    \caption{Gate susceptibility contributions at dispersive frequency $\omega_d=t/5$. We compare Hamiltonian $\chi_{H}$, Sisyphus $\chi_\Gamma$, Hermes $\chi_L$ and quantum capacitance $C_Q$ in imaginary and real part for \textbf{(a)} the ground state at the flux sweetspot, \textbf{(b)} the first excited state at flux sweetspot, \textbf{(c)} the ground state away from the flux sweetspot and \textbf{(d)} the first excited state away from the flux sweetspot. These plots are using the same parameters as Fig.~\ref{fig:KJJ_gate}.}
    \label{fig:KJJ_gate_contributions}
\end{figure}

In Fig.~\ref{fig:KJJ_gate} of the main text, we compare the total common gate susceptibilities at and away from the flux sweet spot for both the ground and first excited state. 
However, each of these susceptibilities consists of several contributions, as discussed in Sec.~\ref{sec:Linear Response in open Systems} of the main text. 
In Fig.~\ref{fig:KJJ_gate_contributions}, we compare each contribution. Note that using the common gate $\mu$ as the perturbed parameter reduces the static susceptibility $\chi_\textbf{st}=0$ and will thus be omitted from the figures. 
For the chosen parameters, Sisyphus contributions are also negligibly small. 

Fig.~\ref{fig:KJJ_gate_contributions}a shows the ground state at the sweet spot. The real part reproduces the quantum capacitance up to a small discrepancy due to thermalization terms of the Hermes contribution. 
The first excited state in Fig.~\ref{fig:KJJ_gate_contributions}b only produces a tiny response. Away from the sweet spot, in Figs.~\ref{fig:KJJ_gate_contributions}c and \ref{fig:KJJ_gate_contributions}d, Hermes and Hamiltonian parts combine the qubit response in a double peak structure. The differences to the quantum capacitance are enhanced because the energy spectrum opens a gap between the lowest energy states. 

\begin{figure}
    \centering
    \includegraphics[width=0.3\linewidth]{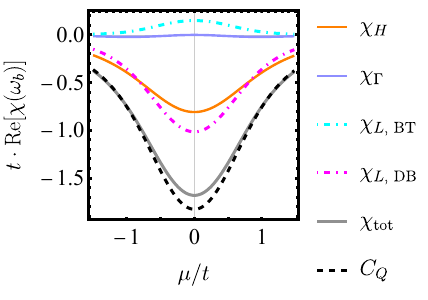}
    \caption{Real part of the gate susceptibility shown in Fig.~\ref{fig:KJJ_gate_contributions}a. We have separated the Hermes contributions $\chi_L=\chi_{L, \text{BT}}+\chi_{L, \text{BT}}$ into contributions of detailed balance (DB) and beyond thermalization (BT). Adding up $\chi_H+\chi_{L, \text{DB}}$ perfectly reproduces quantum capacitance $C_Q$ for dispersive frequencies $\omega_d$.}
    \label{fig:KJJ_gate_contributions_BT}
\end{figure}

In Fig.~\ref{fig:KJJ_gate_contributions_BT} we exemplify the impact of Hermes contributions beyond thermalized states.
We separate the total Hermes susceptibility $\chi_L(\omega)$ in Eq.~\eqref{eq:HermesFinalAppendix} into two contributions according to the factor $\Lambda_{mn} = - (p_m - p_n ) \Gamma_{T_2}^{mn} + \mathcal{G}_m - \mathcal{G}_n$ in Eq.~\eqref{eq:LambdaCoefficientsHermes}.
We label the first term as "detailed balance" (DB) and the second term as "beyond thermalization" (BT) contributions.
Interestingly, without $\chi_{L,\text{BT}}$ and $\chi_\Gamma$, the real part of susceptibility would perfectly reproduce the quantum capacitance. However, BT contributions dampen the signal slightly. This effect is even more prominent when the lower energy bands interact directly, thus away from the flux sweet spot. Therefore, $\chi_{L,\text{BT}}$ is generally not negligible when comparing dispersive measurements.

\subsection{KJJ contributions in flux susceptibility}
\label{Appendix:Susceptibility_contributions_KJJ_flux}
\begin{figure}
    \centering
    \includegraphics[width=\linewidth]{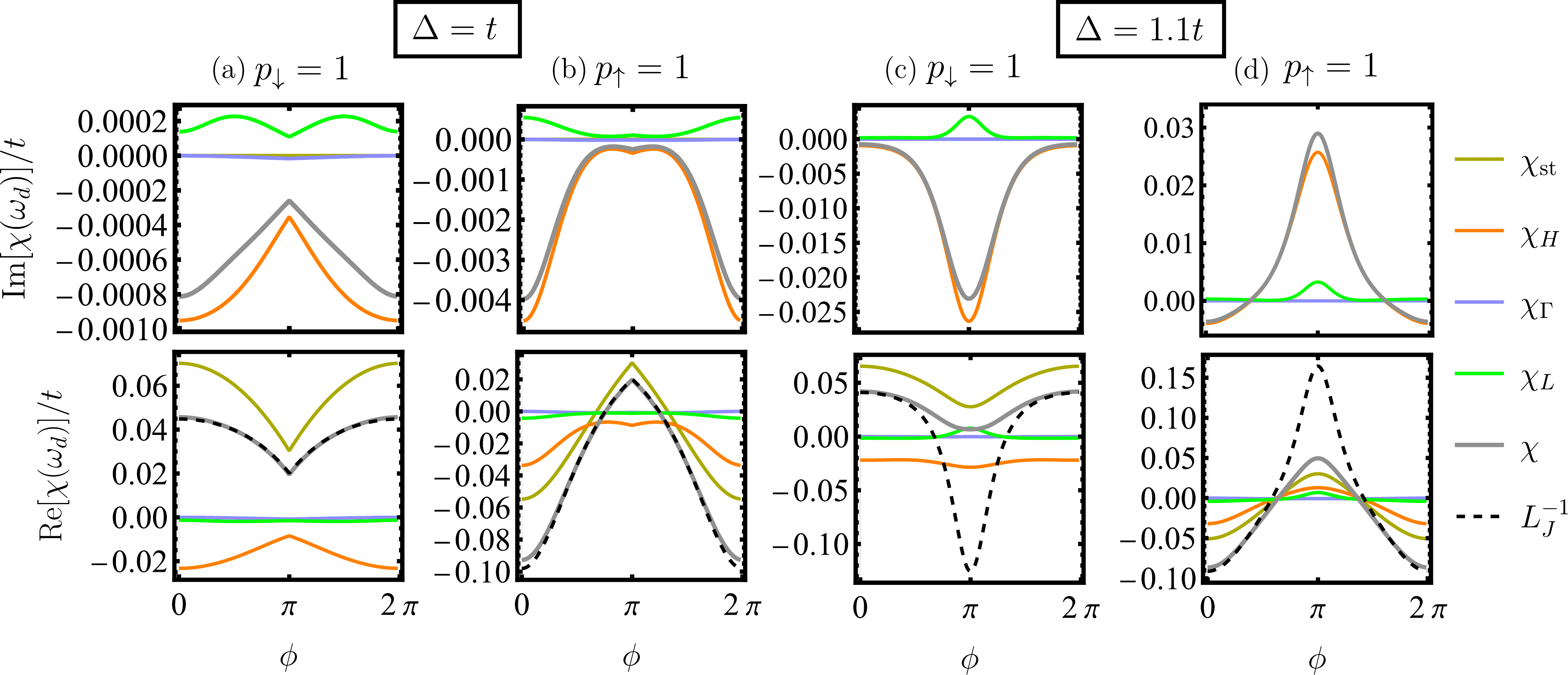}
    \caption{Flux susceptibility contributions at dispersive frequency $\omega_d=t/5$. We compare static $\chi_\text{st}$, Hamiltonian $\chi_{H}$, Sisyphus $\chi_\Gamma$, Hermes $\chi_L$ and curvature $E_n^{\prime\prime}$ terms in imaginary and real part for \textbf{(a)} the ground state at the coupling sweetspot, \textbf{(b)} the first excited state at coupling sweetspot, \textbf{(c)} the ground state away from the coupling sweetspot and \textbf{(d)} the first excited state away from the coupling sweetspot. These plots are using the same parameters as Fig.~\ref{fig:KJJ_flux}.}
    \label{fig:KJJ_flux_contributions}
\end{figure}

The flux susceptibility in Fig.~\ref{fig:KJJ_flux} can be understood by separation into the contributions shown in Fig.~\ref{fig:KJJ_flux_contributions}. 
Importantly, the energy spectrum is not linearly dependent on $\phi$, resulting in $\chi_\text{st}\neq0$. For the chosen parameters, Sisyphus contributions are negligible. At the coupling sweet spot, the imaginary parts nearly vanish as all resonant transitions are at much higher energies than $\omega_d$. The contributions of the real part are dominated by static and Hamiltonian contributions, adding up to the curvature of the bands. Only in Fig.~\ref{fig:KJJ_flux_contributions}b we can see a small difference to the flux curvature because of beyond thermalization terms $\chi_{L,\text{BT}}$ in the Hermes contribution. Away from the coupling sweet spot, $\Delta\neq t$, the static, Hamiltonian, and Hermes susceptibility show peaks around $\phi=\pi$. However, flux band curvature predicts these peaks to be much more prominent. In particular, Eq.~\eqref{eq:dispersiveLimitHamiltonianAndHermesAppendix} shows that the static $\chi_\text{st}$, Hamiltonian $\chi_H$, and beyond thermalization terms of Hermes $\chi_{L,\text{BT}}$ dampen these peaks significantly below the magnitude predicted by the band curvature.

\subsection{Common gate spectroscopy on left chain}
\label{Appendix:Left_chain}

\begin{figure}
    \centering
    \includegraphics[width=.5\linewidth]{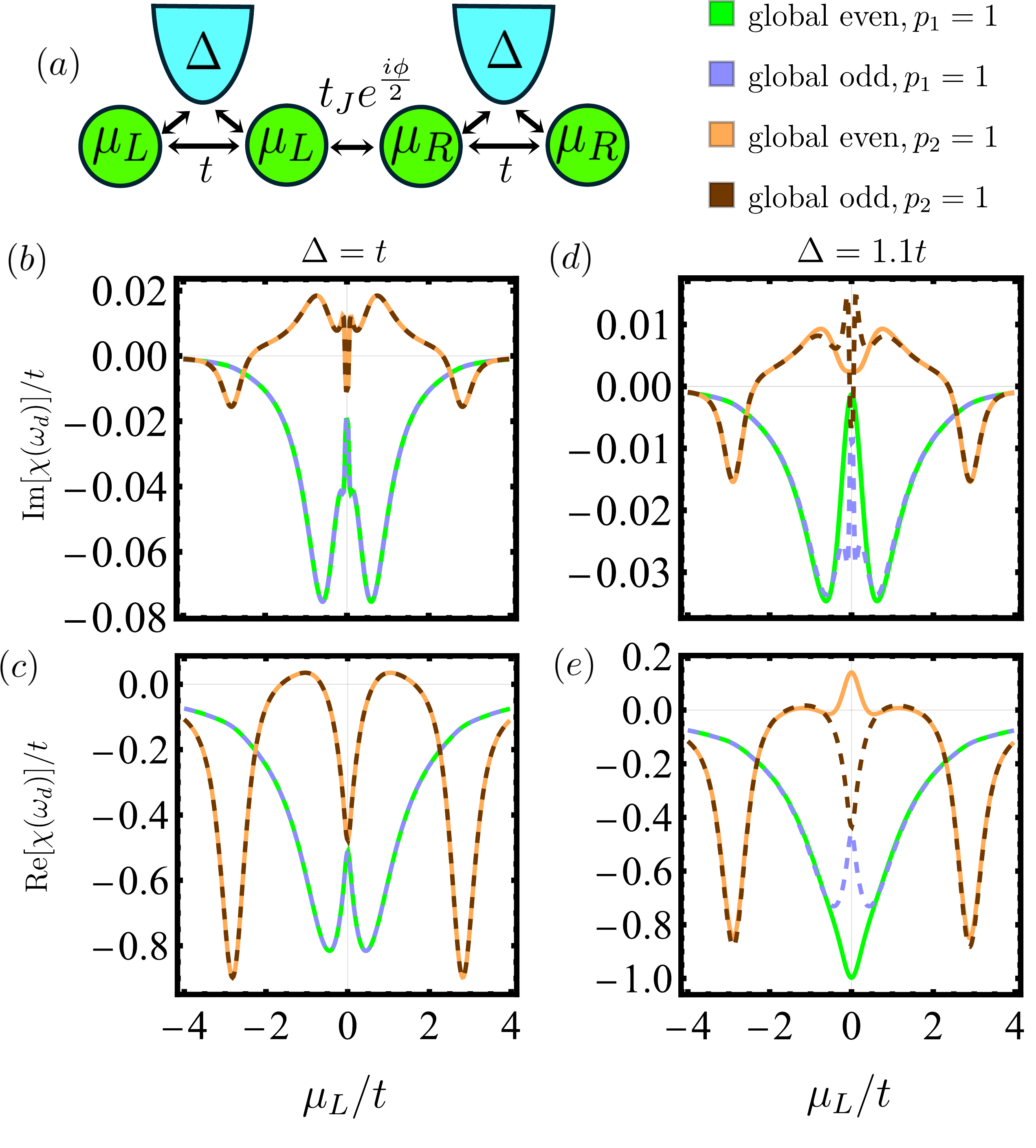}
    \caption{Dispersive susceptibility measurements driving only the left chain common gates.
    \textbf{(a)} Sketch of the KJJ distinguishing left (right) common onsite potentials $\mu_{L(R)}$.  
    \textbf{(b)} Imaginary and \textbf{(c)} Real part of total susceptiblity in global odd and global even parity for the lowest and second lowest eigenstates at the coupling sweetspot.  
    \textbf{(b)} Imaginary and \textbf{(c)} Real part of total susceptiblity in global odd and global even parity for the lowest and second lowest eigenstates away from the the coupling sweetspot.
    Parameters: $ k_BT=\Delta/10, \phi=\pi,|s_{mn}|^2=1/10, t_J=t/2,\mu_R=0$}
    \label{fig:KJJ_gate_left_chain}
\end{figure}

Perturbing the average onsite energy has a fundamental problem. In experiments, all dots need to be tuned individually by local gates. Therefore, threading a common fluctuation through all the intricacies of such a device will not be a preferable option. However, experiments on single shot parity readout of a minimal Kitaev chain \cite{van_loo_single-shot_2026} have shown the possibility of introducing a voltage fluctuation into the common superconducting lead and therefore driving both dot levels simultaneously. We adopt such an approach and calculate the left common gate susceptibility on the KJJ by only fluctuating $\mu_L$, see Fig.~\ref{fig:KJJ_gate_left_chain}a. At the coupling sweet spot $t = \Delta$, the total susceptibility is depicted in Fig.~\ref{fig:KJJ_gate_left_chain}b and c. Global odd and even parity are indistinguishable by only measuring the left chain. This observation holds true, even when detuning the flux. The side peaks for $p_2=1$ emerge because the second and third energy eigenstates show an avoided crossing. Only when detuning $\Delta/t$, we can distinguish in Fig.~\ref{fig:KJJ_gate_left_chain}d and e between global parities close to $\mu_L=0$. A global odd signal is unchanged after detuning the couplings while, for global even parity, new peaks emerge. Therefore, intuition for parity measurement is similar to the flux spectroscopy. Only when tuning $\Delta\neq t$ we can distinguish global odd and global even parity states.

\twocolumngrid
	
	
	\bibliographystyle{apsrev4-2}
	\bibliography{LindbladLinResponsePaper.bib}

\end{document}